\RequirePackage{fix-cm}
\documentclass[12pt]{article}   
\usepackage{graphicx}

 \usepackage{mathptmx}      
\usepackage{tikz}
\usepackage{amssymb}
\usepackage{amsmath, amsthm,enumitem}
\usepackage{subcaption,ulem}

\newcommand{\ord}{ord}
\usepackage{color}

\newcommand{\Var}{Var}
\newcommand{\Cov}{Cov}
\newcommand{\Exprand}{\mathrm{Exp}}
\newcommand{\Ncyc}{N_{\mathrm{c}}}
\newcommand{\dTcyc}{\Delta T_{\mathrm{c}}}
\newcommand{\dTfull}{\dTcyc|\mathcal{A}}
\newcommand{\dTattb}{\Delta T_{1,2}}
\newcommand{\dTdet}[1]{\Delta T_{#1}}
\newcommand{\dMcyc}{\Delta M_{\mathrm{c}}}
\newcommand{\dMfull}{\dMcyc|\mathcal A}
\newcommand{\dMattb}{\Delta M_{1,2}}
\newcommand{\dMdet}[1]{\Delta M_{#1}}
\newcommand{\dMdetjump}[1]{\Delta M^{(\mathrm{d})}_{#1}}
\newcommand{\dMattjump}[1]{\Delta M^{(\mathrm{a})}_{#1}}
\newcommand{\dMattjumpall}[1]{\widetilde{\Delta M}^{(\mathrm{a})}_{#1}}
\newcommand{\dMTfull}{\Delta M, \Delta T|\mathcal A}

\newcommand{\Talldet}{T}
\newcommand{\Malldet}{L}
\newcommand{\dettime}{\tau_{\mathrm{d}}}
\newcommand{\atttime}{\tau_{\mathrm{a}}}
\newcommand{\bAtstate}{\mathbf{Q}}
\newcommand{\batstate}{\mathbf{q}}
\newcommand{\Atstate}[1]{Q^{(#1)}}
\newcommand{\atstate}[1]{q^{(#1)}}
\newcommand{\Rdet}{R^{(\mathrm{d})}}
\newcommand{\detmot}{J^{(d)}}

\newcommand{\attmot}{J^{(a)}}

\newcommand{\detmotfull}{ \detmot{}|\mathcal A}

\newcommand{\attmots}{\varpi}
\newcommand{\pdet}[1]{p_{d}^{(#1)}}
\newcommand{\pdeta}[1]{ p_{d|\mathcal A}^{(#1)}}

\newcommand{\dMstate}[1]{\Delta M_{#1}}
\newcommand{\Vstate}[1]{V_{#1}}
\newcommand{\Dstate}[1]{D_{#1}}
\newcommand{\Tstate}[1]{T_{#1}}
\newcommand{\Ydetb}[1]{Y^{(i)}_{\mathrm{d}}}

\newcommand{\Yatt}[1]{Y^{(#1)}_{\mathrm{a}}}

\newcommand{\barvtwo}{\Vstate{1,2}}
\newcommand{\detone}[1]{\bar d^{\ast}_{#1}}
\newcommand{\dettwo}[1]{\bar d^{\ast(#1)}_{1,2}}
\newcommand{\dettwoboth}{\bar d^{\ast}_{1,2}}
\newcommand{\dettwoz}[1]{\bar d^{(#1)}_{1,2}}
\newcommand{\E}{\mathbb E}

\newcommand{\Rtimegen}{T}
\newcommand{\Rlengthgen}{L}
\newcommand{\Rlength}[1]{\Rlengthgen^{(#1)}}

\newcommand{\Rtime}[1]{\Rtimegen^{(#1)}}
\newcommand{\Nexp}{S}
\newcommand{\Vpool}{V_{\mathrm{ens}}}
\newcommand{\Vrun}{V_{\mathrm{run}}}
\newcommand{\Dpool}{D_{\mathrm{ens}}}
\newcommand{\Drun}{D_{\mathrm{run}}}

\newcommand{\totdetprob}{p^{\emptyset}_d}
\newcommand{\dTstate}[1]{\Delta T_{#1}}
\newcommand{\RdetJ}[1]{R^{(d)}_{#1}}

\newtheorem{proposition}{Proposition}
\newtheorem{corollary}{Corollary}

\begin{document}

\title{Effective behavior of cooperative and nonidentical molecular motors
}

\date{}

\author{Joseph J. Klobusicky        \and John Fricks \and 
        Peter R. Kramer 
}



\maketitle

\begin{abstract}
Analytical formulas for effective drift, diffusivity, run times, and run
lengths are derived for an intracellular transport system consisting of a
cargo attached to two cooperative
but not identical molecular motors (for example, kinesin-1 and kinesin-2)
which can each attach and detach from a microtubule.  The dynamics of the
motor and cargo in each phase are governed by stochastic differential equations,
and the switching rates depend on the spatial configuration of the motor
and cargo.  This system is analyzed in a limit where the detached motors
have faster dynamics than the cargo, which in turn has faster dynamics than
the attached motors.  The attachment and detachment rates are also taken
to be slow relative to the spatial dynamics.  Through an application of iterated
stochastic averaging to this system, and the use of renewal-reward theory
to stitch together the progress within each switching phase, we obtain explicit
analytical expressions for the effective drift, diffusivity, and processivity
of the motor-cargo system.  Our approach accounts in particular for jumps
in motor-cargo position that occur during attachment and detachment events,
as the cargo tracking variable makes a rapid adjustment due to the averaged
fast
scales.  The asymptotic formulas are in generally good agreement with direct
stochastic simulations of the detailed model based on experimental parameters
for various pairings of kinesin-1 and kinesin-2 under assisting, hindering,
or no load.
\end{abstract}

\emph{Dedicated to Andy Majda for his 70th birthday, with gratitude for his
lasting inspiration starting from my undergraduate and graduate days on the
creative deployment of mathematical modeling and the beautiful application
of analysis techniques as a lens for exploring and understanding the dynamics
of physical systems - PRK}

\section{Introduction}
\label{intro}

A   biological cell  during its interphase requires  sufficiently fast transport
of organelles and other compounds for its survival   \cite{alberts2017molecular}.
  Transport through diffusion alone is often far too slow. To illustrate,
a compound moving through pure diffusion in  some neurons might take years
to travel over the cell's length \cite{phillips2012physical}.      For eukaryotic
organisms, intracellular trafficking of vesicles is instead governed by directed
transport along a network of thin filaments, such as microtubules or actin.
 A vesicle and the molecular compound it encloses, collectively referred
to
as a cargo, travel along the filaments by attaching to one or several molecular
motors.  As an important example on which we will focus, we can consider
molecular motors called kinesins, which consist of
two heads which attach to a microtubule, a tail which attaches to the cargo,
and a coiled-coil tether connecting the heads and tail \cite{hancock2003kinesins}.
 But the mathematical framework to be developed can be applied to more general
molecular motors, including dynein and myosin.

%

The motor-cargo attachment is generally found to be much more durable than
the motor-microtubule attachment~\cite{hancock2014bidirectional},  so models
of motor-cargo complexes typically
assume the number of motors attached to a given cargo can be treated as a
fixed constant $ N $ over the transport time scale of interest~\cite{WangLi:2009,kunwar2011mechanical,muller2008tug,mallik2013teamwork,lombardo2017myosin}.
But the
number of those $ N$ motors that are attached to  microtubules and therefore
actively engaged in transport does appear to fluctuate through dynamical
attachment and detachment of the motors to and from the microtubule.  In
the present work, we contemplate the simplest scenario in which the motor-cargo
complex is in the vicinity of a single microtubule.  
The state of the motor-cargo complex can then be classified in terms of which
of its motors  
are attached to the microtubule, and therefore engaged in directed motion.
 Our model could also be formally applicable for a bundle of parallel microtubules
with common polarity if they are spaced sufficiently closely that the progress
of the cargo is not so sensitive to what particular set of microtubules the
motors are attached. Experimental observations \cite[Fig. S5]{rai2013molecular}
show
that approximating the multiple motors as all attached to a single microtubule
could be consistent even for some situations in cell.
 
\begin{figure}
\includegraphics[width = \textwidth]{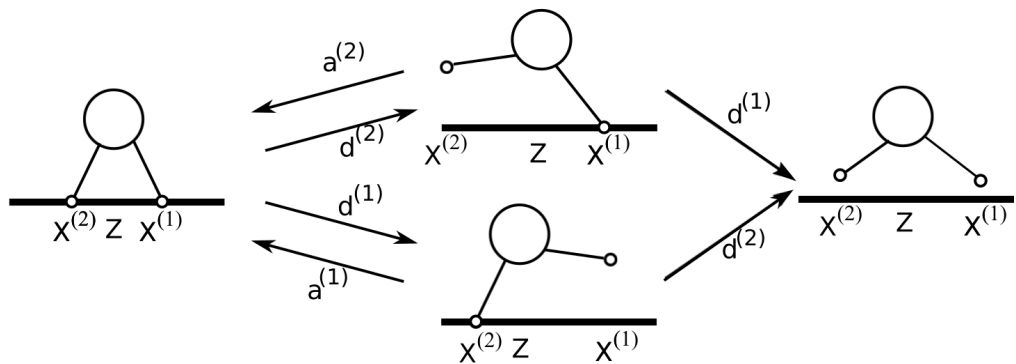}
\caption{Attachment and detachment from a microtubule for a system of two
motors durably attached to a cargo.  
The spatial
positions along the microtubule for the motors are denoted $ X^{(1)}$, and
$
X^{(2)} $, and for the cargo by $ Z$. From a state with two attached motors
(left),
each motor $i $ can detach with a rate $ d^{(i)} $ depending on the current
spatial
configuration.  From a state with one attached and one detached motor (center),
the detached motor $i$ can (re)attach with a constant rate $ a^{(i)} $, or
the
attached motor $ i^{\prime} $ can also detach at a configuration-dependent
rate $ d^{(i^{\prime})}  $, terminating the run on the microtubule.
}
\label{fig:switch}       
\end{figure}

A cargo with two motors, for example, can fluctuate between three possible
states while remaining connected to the microtubule:\ two states with one
attached and one detached motor, and one state with two attached motors (Figure~\ref{fig:switch}).
 One further state has both motors detached, after which we will consider
the cargo to move away from the microtubule and terminate its run, though
one could contemplate the motors remaining weakly bound and possibly sliding
along the microtubule~\cite{smith2018assessing,muller2008tug}.
 
Microtubules are oriented with a + and - end, with the molecular motor kinesin
always traveling
from the - to + end \cite{alberts2017molecular} and the molecular motor dynein
traveling in the
opposite direction.  This can result in ``tug
of war'' scenarios and bidirectional transport for ensembles including such
antagonistic
motors~\cite{muller2008motility}.
 Our focus here is on \emph{cooperative} transport of a cargo by possibly
different motor types which each  move, when considered in a single motor
complex, in the same direction on the microtubule, such
as two different types of motors from the kinesin superfamily.  
Since we wish to contemplate
the possibility of the motors being of different types, motor
labelling is relevant and we distinguish between the two states with one
motor attached.  This is in contrast to models with $N $ identical motors
attached, which can be classified more compactly in terms of simply the number
of those motors which are currently attached to a microtubule \cite{klumpp2005cooperative}.
For multiple motor complexes, we are  interested in calculating how
statistics such as run lengths and effective speeds depend on the properties
of the individual motors.
   Comparisons of statistics between  motor complexes are not always intuitive.
For instance, it has been observed in~\cite{feng2018motor} that heterogeneous
motor complexes
of kinesin-1 and kinesin-2 have longer run lengths than pure systems
consisting
only of kinesin-1.  This observation would seem to depend upon a variety
of physical
parameters, often interacting in complex ways. Indeed, while  kinesin-2 is
about half as fast as kinesin-1 and detaches more readily under load, it
also appears to reattach to microtubules four times more quickly than kinesin-1~\cite{feng2018motor}.

A number of properties of some individual motors, acting in isolation, have
been obtained from in vitro experiments with optical traps, in
which a polystyrene bead, serving as a cargo, is attached to a motor, and
directed along a microtubule with an applied optical trap force.    Under
this
setting, several motor properties can be measured, particularly their speed,
diffusivity, and detachment rate as a function of the applied optical trap
force \cite{spudich2011optical,milic2014kinesin,milic2017intraflagellar,andreasson2015mechanochemical,leduc2004cooperative,block1990bead,vale1996direct}.
 The above experimental work can be used as a basis for parameterizing biophysically
mechanistic models for individual motors in theoretical models exploring
their interactions~\cite{kunwar2011mechanical,kunwar2010robust,lipowsky2009review,klumpp2015review,lipowsky2013network,bouzat2016models,Jamison:2010,kolomeisky2016collective,srinivas2019kramers,driver2010coupling,miles2018analysis}.
For the purpose of experimentally measuring the interaction of molecular
motors, and in particular for motivation for and comparison against theoretical
models, an important  experimental development has been to use DNA origami
for cargo, which specifies closely arranged handle sites onto which specified
motor types attach~\cite{feng2018motor,Jamison:2010,furuta2013measuring}.
 The actual engagement of motors with microtubules
cannot be resolved, so particularly when not using engineered motor-attachment
constructs, the number of relevant motors of various types attached to an
observed cargo with a particular microtubule must typically be statistically
inferred, sometimes using simple theoretical models~\cite{rai2013molecular,encalada2011assemblies,lombardo2017myosin}.

\subsection{Overview of Methods and Results}

Our mathematical modeling framework is based on the one
developed in McKinley, Athreya, et al~\cite{mckinley2012asymptotic}, where
cargo transport statistics were examined
for
cooperative ensembles of  identical motors that are treated as permanently
attached to the microtubule.   The coupled spatial dynamics of the motor
and cargo position are expressed as a system of continuous stochastic differential
equations (SDEs) which are viewed as a coarse-graining of discrete stepping
model.  The model is posed in essentially one spatial dimension along the
microtubule, neglecting transverse fluctuations.  We extend the model  in~\cite{mckinley2012asymptotic}
in two ways:  1) allowing
the motors to be distinct, but still cooperative, and 2) allowing the motors
to attach and detach from the microtubule.  This model represents the motor-cargo
system in terms of basic  parameters concerning biophysical properties of
the individual motors and the cargo.  The
cargo and detached motor dynamics are modeled as overdamped point particles
responding to spring forces from the motor-cargo tether and driven by stochastic
terms representing thermal fluctuations.  On the other hand, as in~\cite{mckinley2012asymptotic},
the attached motors dynamics are governed by a nonlinear force-velocity relation
together with stochastic terms arising from the nonequilibrium stepping process.
Punctuating the continuous evolution of the motor-cargo dynamics is the attachment
or detachment of motors from the microtubule.  These are represented in terms
of Markovian jump processes, with attachments occurring at constant rates
but the detachments occurring at rates dependent upon the force on the motor.

To characterize the effective behavior of the motor-cargo
complex as a whole, we proceed through a sequence of coarse-graining steps
motivated by the separation of time scales of the various physical processes.
 First we average over the fast dynamics of the cargo and detached motors.
  This can be accomplished easily for arbitrary number $N $ of cooperative
motors, but our further analytical progress is restricted to the case of
$N=2 $ cooperative motors due to the substantial increase in complexity for
$ N> 2$.  We homogenize over the spatial dynamics of the attached motors
between attachment and detachment events.  This eventually yields an effective
Markov chain description in terms of jumps between different states of motor
attachment, with associated durations and displacements of the motor-cargo
complex within each state.  By decomposing the stochastic trajectory through
this finite state space in terms of cycles demarcated by moments when both
motors are attached, we further coarse-grain the motor-cargo dynamics into
a renewal-reward process.  With Wald's identity, we characterize the fundamental
statistics of the run length and run time of the motor-cargo complex in terms
of the statistics of duration and displacement over one of these cycles,
which are in turn related to the
basic biophysical parameters of the model.  With 
 a law of large numbers argument and its application to renewal-reward processes,
we relate the effective velocity and diffusivity also to the cycle statistics,
and thereby in turn to the basic biophysical parameters.   We next proceed
to discuss the components of our analysis in some more detail, with reference
to related work.

\subsection{Modeling approach}

 The reason we work with the coarse-grained SDE model for the motors is to
reduce the number of physical parameters modeling the individual motor to
those characterizing the force-velocity relationship, the force-detachment
relationship, and a noise parameter.  Moreover, the mathematical presentation
is simplified by having the cargo and motor dynamics in a unified SDE framework.
 One could of course start with a motor stepping model~\cite{bouzat2016models,reld:rgovm,reld:rsmp,hughes2012kinesins,WangLi:2009,tce:mdbt,abk:mm,driver2010coupling,lipowsky2013network},
and readily coarse-grain
it to obtain the parameters for our SDE model, and the conclusions would
be equivalent unless the cargo fluctuations fed back on the force-dependent
kinetics of stepping in a substantially different way than they do on the
force-dependent  coarse-grained velocity.  We cannot rule such subtle feedback
out~\cite{hendricks2009collective}, but if it were real, one would presumably
have
to represent  the chemomechanical stepping cycle in more detail to capture
these
effects than a generic stepping model with one step per cycle.  The detailed
description of the stepping dynamics of even well-studied
motors such as kinesin-1
is still under active experimental investigation~\cite{andreasson2015mechanochemical,hancock2017vulnerable,milic2014kinesin},
while
the more coarse-grained characterizations needed for the SDE description
are more consistently established across labs.  For these reasons, we will
proceed here with the coarse-grained SDE description, in somewhat the same
vein as coarse-grained integrate-and-fire models are often used in place
of more detailed Hodgkin-Huxley models to study networks of interacting neurons
\cite{vanVreeswijk:1993,young2015emergent,gk:fpdcb,JJHopfield07181995,OBH09,Shkarayev:2012,zillmer:031909,kan:dcbpd}.
 
In Section~\ref{sec:model}, we present our extended model
allowing for distinct motors and attachment/detachment from the microtubule.
The modification of the spatial
dynamics is presented in Subsection~\ref{ssec:driftmodel}, resulting in a
system of  SDEs (\ref{motors})-(\ref{driftu}) with the motor dynamics depending
on both the motor label and whether the motor is attached or detached.  In
Subsection~\ref{ssec:switchsect},
we present our model of switching times. Attachment rates are rather
difficult to quantify experimentally and appears to depend on the operating
conditions and whether the motor is tethered near the microtubule by other
motors~\cite{Jamison:2010,furuta2013measuring,mallik2018coin},
and we simply adopt the common approach of modeling them as constant (as
in a homogenous Poisson process model).  The detachment rates, on the other
hand, will be taken as functions of the applied force on the attached motor,
which is itself a stochastic process induced by the stochastic spatial dynamics.
 Mathematically, then, the detachment process is more akin to a Cox process
\cite{cox1980point,cox1955some}.
The class of 
functions we consider for detachment rates is general enough to  include
the most common cases seen in previous works,
including constant \cite{baker2000thermodynamics}, exponential \cite{andreasson2015mechanochemical,kunwar2010robust,kunwar2008stepping,klumpp2005cooperative},
double-exponential functions~\cite{milic2017intraflagellar}, and exponential
functions turning over
to slower linear growth beyond stall~\cite{kunwar2011mechanical,vershinin2018intersect}
or along the assisting direction~\cite{arpag2019kin1kin3}.

Transitions between attached and detached  states
have been modeled as continuous time  Markov chains in the case of identical
motors \cite{klumpp2005cooperative,materassi2013exact,talukdar2016steady,klumpp2015review}
with constant  switching rates depending
on the number of attached and detached motors. Our model allows for nonidentical
motors, which increases the state space of possible attachment
and detachment
configurations.  Similar to the discrete-space models discussed in  \cite{materassi2013exact,talukdar2016steady},
our transition rates depend on relative motor positions, but now must consider
motor types.  Moreover, our model does not make the popular assumption~\cite{klumpp2005cooperative,klumpp2015review,Jamison:2010}
that the cargo is always in mechanical equilibrium with
the motors nor the further mean-field approximation that all motors feel
a perfectly shared load from the cargo.  Stochastic fluctuations of cargo
and motor
positions have been shown to significantly alter the mean-field predictions,
at least for small teams of motors~\cite{kunwar2010robust,uppulury2013varying,kunwar2011mechanical},
and experimental observations do not support the
load-sharing assumption in teams of two kinesin-1~\cite{Jamison:2010}.  
Bouzat~\cite{bouzat2016models} shows how the neglect of fluctuations in the
cargo position can give incorrect modeling results for the effective force-velocity
response of a single motor and in particular for the stall force of cooperative
motors.  Berger, M\"{u}ller, and Lipowsky~\cite{lipowsky2009myosin} and Arpa\u{g},
Norris, et al~\cite{arpag2019kin1kin3} similarly
find
 the force felt by a  motor from the cargo substantially affects its
effective detachment rate, and so departs from the mean-field dynamical description
by averaging the detachment rate against the modeled probability distribution
of the force applied by the cargo in a given configuration of attached and
detached motors.  We apply a similar, but more systematic, procedure in Subsection~\ref{subsub:effdet}
to compute effective detachment rates by averaging over a quasi-stationary
distribution of the force felt from the cargo, as a function of the current
configuration of attached motors.
Uppulury, Efremov, et al~\cite{uppulury2013varying} and Kunwar, Tripathy,
et al~\cite{kunwar2011mechanical}
 conducted informative simulation studies of discrete-state
stochastic stepping models that keep track of the relative positions of identical
motors, but did not develop analytical formulas relating  collective behavior
to single-motor properties.  Keller, Berger, et al~\cite{lipowsky2013network}
similarly conducted
simulation studies for a pair of identical motors which resolved the chemomechanical
steps, finding in particular that the transport is degraded as the strength
of the Hookean tether coupling the motors and cargo is increased.
Wang and Li~\cite{WangLi:2009} derive an analytical formula for the effective
velocity
of a cooperative team of identical stochastic stepping motors, but don't
include attachment and detachment effects.  Li, Lipowsky, and Kierfeld~\cite{lipowsky2013bifurcation}
developed a variation of the Markov chain model of Klumpp and Lipowsky~\cite{klumpp2005cooperative}
for cooperation by two groups of motors, one fast and one slow, in the context
of gliding assays where the numbers of engaged motors are considerably larger
than for cargo transport, to relate bistability in the transport of microtubules
to a small ratio of detachment force scales to the stall forces.
A recent model by Miles, Lawley, and Keener~\cite{miles2018analysis} for
cooperative molecular motor
transport employs a similar framework to ours, with a stochastic differential
equation
for the cargo dynamics, but with step-resolving dynamics for the motors.
  The model we present extends the model of~\cite{miles2018analysis} in allowing
the
motors to be nonidentical and allowing the detachment rates of motors to
depend on the spatial configuration of the motors.    

\subsection{Computation of transport statistics} 
The analysis of our model will proceed in Section~\ref{sec:nond} by an extension
of
the asymptotic analysis developed in McKinley, Athreya, et al~\cite{mckinley2012asymptotic},
based
on the cargo dynamics being taken as fast relative to the dynamics of the
motors attached to the microtubule.  Extending this nondimensionalization
to our setting of nonidentical motors which attach and detach from the microtubule
in Section \ref{sec:nondim}, we further motivate taking the detached motor
dynamics to be faster than the cargo (due to their relative size), and the
attachment and detachment processes as slow compared to the dynamical time
scale of attached motors.  A similar hierarchy of time scales was considered
in the context of the stepping of the two heads of a kinesin motor attached
to a cargo in Peskin and Oster~\cite{Peskin:1995}, where the detached head
was taken to move
quickly relative to the cargo, while the cargo dynamics were taken as fast
relative to the time scale of motor stepping.

We set up an asymptotic analysis
which formalizes the separation of time scales we identify,
with the further assumption that the parameters
for different motor types do not vary drastically (which seems to be reasonable
for kinesin-1 versus kinesin-2, for example,~\cite{andreasson2015mechanochemical}).
 We proceed in Subsection~\ref{sec:mavg} to apply stochastic averaging successively
over the unbound motor positions and the cargo position to obtain effective
dynamics and effective detachment rates based only on the positions and identities
of the attached motors.    The analysis thus far applies to $ N $ cooperative
but possibly not identical motors, and in fact does not require a time scale
separation assumption between the attached motor dynamics and attachment/detachment
dynamics.  In Section~\ref{slowswitch}, we use this scale separation assumption
for the case of $
N=2 $ motors to proceed further by homogenizing the effective
dynamics obtained from the stochastically averaged equations within each
interval between attachment or detachment, and averaging the detachment rates
over the spatial configurations of the motor-cargo complex.

Finally, in Section~\ref{sec:renewal} we develop theoretical formulas for
the effective velocity, diffusivity, and processivity statistics of the motor-cargo
complex, 
all expressed in terms of explicit formulas involving the parameters governing
the motor and cargo dynamics.  
 The appropriate definition of velocity and diffusivity requires some consideration
when applied to random finite time intervals over which the cargo remains
associated to a microtubule, and we consider two versions relevant to different
protocols of computational simulation or experimental observations where
data from multiple runs are collected.  Our method of deriving these effective
transport statistics is through the application of the law of large numbers
and renewal-reward
asymptotics to the progress made by the motor-cargo complex over several
cycles
of attachment and detachment events.  
Krishnan and Epureanu~\cite{krishnan2011renewal} developed some
early ideas for how to interpret experimental statistics for molecular motor
systems using results from renewal-reward theory. Hughes, Hancock, and Fricks~\cite{hughes2012kinesins}
appealed to  renewal theory arguments for a discrete semi-Markov
 stepping model to compute effective statistics for motor head stepping from
a chemomechanical cycle model, and Shtylla and Keener~\cite{shtylla2015burnt}
 applied this
framework
to describe the effective dynamics of a ParB protein complex that moves along
a track of ParA proteins in a manner similar to molecular motors which ``burn''
the track as they progress.
  In a broad sense, our strategy of combining
asymptotic arguments regarding separation of time scales and renewal-reward
calculations to obtain these explicit formulas mirrors that of Miles, Lawley,
and Keener~\cite{miles2018analysis}
for the case of identical cooperative motors with force-independent
detachment rates, though our implementation of these ideas differ in some
details
which we will discuss in the conclusion (Section~\ref{sec:conc}).

We show in Section~\ref{sec:sims} that our theoretical formulas obtained
by these scale separation arguments for the effective transport of kinesin-1/kinesin-1,
kinesin-2/kinesin-2, and kinesin-1/kinesin-2 complexes compare generally
well with direct numerical simulations of our model as parameterized by experimental
observations of individual kinesin-1 and kinesin-2 motors.   A few issues
in translating experimental observations into model parameterization, however,
 require study before our predictions based on our mathematical framework
can meaningfully be compared directly with experimental observations~\cite{feng2018motor,Jamison:2010}
of pair complexes of kinesin-1 and kinesin-2.   Our objective in the present
work is on the development of the general mathematical coarse-graining procedure
and formulas for  cooperative molecular motor transport in terms of the biophysical
properties of the constituent motors, and we provide in Section~\ref{sec:conc}
an assessment of the merits of this mathematical approach as well as limitations
whose resolution will motivate future work.

\section{Models of evolution and switching}\label{sec:model}
For a fixed number of attached and detached motors, we model the transport
of a motor and cargo  system through a system of SDEs. Similar to McKinley,
Athreya, et al~\cite{mckinley2012asymptotic},
we coarse-grain over this discrete stepping and model this motion as a one-dimensional
continuous process along the direction of the microtubule, neglecting transverse
fluctuations which might not be so significant~\cite{driver2010coupling}.
  The locations of  $N$ motors over time $t$  are denoted by
$X^{(i)}(t) \in (-\infty, \infty)$, $i = 1, \dots, N$.  All motors in the
system are assumed to remain
attached to a single cargo, with a position of $Z(t)\in (-\infty, \infty)$.
During a realization
of the stochastic process, a motor will change its state from time to time.
We will denote the $i$th motor's state at time $t$ as $\Atstate{i}(t) \in
\{0,1\}$,
where 0 and 1 denote states of detachment or attachment to the microtubule,
respectively.  The switching
between motor states will be governed by a jump process,  which is Markovian
with respect to the filtration generated jointly by $ \{X^{(i)}(t)\}_{i=1}^N$,
$ Z(t) $ and $ \{\Atstate{i}(t)\}_{i=1}^N$, and 
with switching (jump) rates due to attachment and detachment depending on
the spatial displacements
between the motors and cargo.   

\subsection{Drift and diffusion of nonidentical motors}\label{ssec:driftmodel}
For a motor system with $N$   motors the governing equations take, for $i
= 1, \dots, N$,  the autonomous
form
\begin{align}
dX^{(i)}(t) &= \left(\mu^{(i)}_d(X^{(i)}(t),Z(t))dt+\sqrt{2k_BT/\gamma_{m}^{(i)}}dW^{(i)}(t)\right)\mathbf
(1-\Atstate{i}(t))\nonumber\\&+ \left(\mu^{(i)}_a(X^{(i)}(t),Z(t))dt+\sigma^{(i)}dW^{(i)}(t)\right)
\Atstate{i}(t), \label{motors}
\\
\gamma dZ(t) &= \sum_{j = 1}^N F^{(j)}(X^{(j)}(t)-Z(t)) dt- F_Tdt +\sqrt{2k_BT\gamma
}dW_z(t). \label{cargo}
\end{align}
with  drift coefficients $\mu_a^{(i)}, \mu_d^{(i)} :\mathbb R^2 \rightarrow
\mathbb
R$ for the motors satisfying 
\begin{align}
\mu^{(i)}_a(x,z)&= v^{(i)}g(F^{(i)}(x-z)/F_s^{(i)}),\label{drifta}\\ \mu^{(i)}_d(x,z)&=
-F^{(i)}(x-z)/\gamma_{m}^{(i)}. \label{driftu}
\end{align}
 A table of the parameters and their roles can be found in Table \ref{paramtable}.
  We next briefly summarize the meaning of the model equations (\ref{motors})-(\ref{driftu});
more details can be found in~\cite{mckinley2012asymptotic}. 

If $\Atstate{i}(t) = 1$,  equation (\ref{motors}) describes an attached motor
with
position $X^{(i)}(t). $ The restorative force in (\ref{drifta}) results from
the stretching of the coiled-coil tether that connects   the motor  head
and
tail attached to the cargo.  While a nonlinear force model for the
tether would seem to be most appropriate~\cite{Jamison:2010,uppulury2013varying,vale1997load},
we could not find a clear consensus on its precise form.  To reduce technical
complications in the  formulas and to keep focus on the overall structure
of the mathematical course-graining, in the present work, similarly to Miles,
Lawley, and Keener~\cite{miles2018analysis},
we model the force for this
tether for motor $i$  by a simple Hookean spring relation $F^{(i)}(y) = \kappa^{(i)}
y$,  where $y$ is the longitudinal displacement from the cargo to the
motor, and $\kappa^{(i)}$ is the effective spring constant for the tether
to the $i$th
motor.  The value we have cited from Furuta, Furuta, et al~\cite{furuta2013measuring}
is measured
from motors attached to a DNA scaffold, but we expect the stiffness to mostly
reflect the properties of the motor tether~\cite{driver2010coupling}.  The
value is roughly consistent with the $ \kappa \approx 0.3 $ pN/nm values
found in other experiments~\cite{kojima1997mechanics,kunwar2008stepping,vale1997load}.
 Surely a better model for the tether would have it be slack under compression
from its rest length~\cite{kunwar2008stepping,kunwar2016anisotropy,arpag2019kin1kin3,gross2011arranged,furuta2013measuring}.
  While we do not investigate
the consequences of nonlinear tether force models here in detail, our  framework
can be generalized to include them, as  discussed in Appendix
\ref{app:nonlinspring}.   

The nondimensional force velocity relation
$g:\mathbb R \rightarrow  \mathbb R$ is multiplied by an unencumbered velocity
$v^{(i)}$ to produce an instantaneous expected
velocity.  The argument of $g$ measures the ratio between the tether force
and the motor's stall force $F^{(i)}_s$, or the opposing force needed from
the
cargo to anchor a motor. Positive arguments of $ g $ correspond to forces
opposing or hindering the free motion of the motor.  To agree with the definitions
of $v^{(i)}$ and $F^{(i)}_s$,
$g$ must  satisfy $g(0)= 1$  and  $g(1) = 0$. Random effects are modelled
by independent Brownian motions $W^{(i)}(t)$, with an effective motor diffusion
of $\frac 12(\sigma^{(i)})^2$. In Table \ref{paramtable}, we have used randomness
parameters for kinesin 1 and 2 found in \cite{visscher1999single,andreasson2015mechanochemical}
to calculate diffusivities via their relation which can, for example, be
found in Eqn. (47) of Krishnan and Epureanu~\cite{krishnan2011renewal}.
     
When $\Atstate{i}(t) = 0$, the equation for the position of an detached motor
$X^{(i)}(t)$
in (\ref{motors}) is an overdamped Langevin equation for a  particle with
a friction constant $\gamma_{m}^{(i)}$, and spring constant $\kappa^{(i)}.$
 The friction constant $\gamma_{m}^{(i)}$ was computed with the Stokes-Einstein
relation $\gamma_{m}^{(i)} = 6\pi a \eta $ for a spherical object with
radius
$a=50$ nm in water, with a fluid dynamic viscosity of $\eta =10^{-9}$ pN
s/nm\textsuperscript{2}.  
The
Brownian
motions $W^{(i)}(t)$ are independent of each other, and also of the Brownian
motion $W_z(t)$ driving the cargo. Finally,  the constant $k_BT$ is the Boltzmann
constant multiplied by temperature. 

The equation for cargo position $Z(t)$ in (\ref{cargo}) also follows an overdamped
Langevin equation with friction constant $\gamma$ (also calculated using
the Stokes-Einstein law with $a = 500$ nm), but differs from that
of  detached motors in two ways.  First, the cargo is subject to
spring
forces from each of the $N$ motors. Second, we also account for a possible
constant applied optical trap force $F_T$, as in the experiments of~\cite{kunwar2011mechanical,milic2017intraflagellar,milic2014kinesin}
and simulations of~\cite{bouzat2016models,hendricks2009collective,klumpp2005cooperative,kunwar2010robust,kunwar2016anisotropy,Jamison:2010}.
 Note that the friction constant
values for the detached motors and cargo should be viewed in a somewhat notional
sense, and we have not attempted to give them precise values.  Fortunately
as will discuss in Section~\ref{sec:conc}, the effective dynamics are not
sensitive to their precise values.

\subsection{Switching between attachment configurations}\label{ssec:switchsect}
We model the transition between attachment configurations with varying numbers
of attached
and detached motors with jump processes. See Table \ref{paramtable}
for a list of typical attachment/detachment values.
The attachment of motors is modeled by a homogeneous Poisson process, with
each detached motor having  an attachment rate of   $a^{(i)}$.   As in most
theoretical work~\cite{klumpp2005cooperative,kunwar2011mechanical,bouzat2016models},
we take the attachment rates
to be independent of the configuration of the motor-cargo complex, though
we acknowledge that Furuta, Furuta, et al~\cite{furuta2013measuring} finds
significant reduction
of the attachment
rate of a second motor when the cargo is under load.  The attachment
rate of a motor can naturally be expected to be somewhat different when at
least one other motor on the same cargo is also attached to a microtubule
than when none are~\cite{miles2018analysis}, as does seem to be indicated
experimentally~\cite{Jamison:2010}.
 As we do not model the attachment of the motor-cargo complex from a state
of no attached motors, the attachment rate we require is the one where at
least one other motor on the same cargo is currently attached to a microtubule.
 We therefore take our parameter value for $ a^{(i)} $ in Table~\ref{paramtable}
from recent experimental measurements in this setting~\cite{feng2018motor},
rather than the conventional
estimate in modeling work~\cite{klumpp2005cooperative} based on single-kinesin
attachment rates.  When a motor reattaches, we simply place it at the same
position along the microtubule as it was previously in the detached state.
 
 This is a bit different than most other reattachment models in the literature,
which do not precisely track the motor position in the detached state (as
we do), but apply a selection rule of where the motor reattaches.  For example,
\cite{lipowsky2013network} attaches the second motor at a location that leads
to a zero
tether force with the cargo, while 
\cite{miles2018analysis} attaches the
motor directly at the current cargo position.  By contrast, 
\cite{furuta2013measuring,bouzat2016models,kunwar2016anisotropy,gross2011arranged}
randomly choose attachment
sites that are within a geometrically
defined range of the motor's attachment point to the cargo while~\cite{lombardo2017myosin,driver2010coupling}
preferentially reattaches a motor to the microtubule
at locations that require the least strain energy on the motor.  Our approach,
as well as the other random models just cited, are consistent with experimental
observations that a motor can reattach ahead or behind the motor already
attached~\cite{feng2018motor}.

Detachment rates are determined through the function depending on the force
$ F $ felt by the motor:
\begin{equation}\label{detach}
d^{(i)}(F) = d^{(i)}_0\Upsilon^{(i)}(F/F_d^{(i)}).
\end{equation}
Here, the parameter $F_d^{(i)}$ is a scale force, and   $\Upsilon^{(i)}$
satisfies
$\Upsilon(0)
= 1$, so that $d^{(i)}_0$ is the detachment rate under no external
force.  
Note that since we are only modeling the motor-cargo dynamics along the microtubule
direction, our detachment rate model correspondingly depends only on the
longitudinal force $ F $, with \emph{a positive sign corresponding to opposing
the motor's natural direction of motion.}  Note this convention of writing
the relation of the detachment rate and applied force with positive arguments
corresponding to a hindering force is opposite to how such relations are
typically presented in recent experimental studies~\cite{milic2014kinesin,milic2017intraflagellar,howard2019direction}
 but are consistent with how force-velocity relationships are typically expressed
in theoretical and simulation studies~\cite{klumpp2015review,bouzat2016models,mckinley2012asymptotic}.
The force $ F=F^{(i)} (t) $ opposing the motor $i$
at time $t$ consists  of
the force from the tether to the cargo, which is\begin{equation}
F^{(i)}(t) = \kappa^{(i)}(X^{(i)}(t)-Z(t)). 
\end{equation}
As $ F^{(i)} (t) $ is a random process, so is the detachment rate $d^{(i)}(F^{(i)}
(t) )$ of motor $ i $.  
The resulting model for detachment thus amounts
to a Cox process,  a generalization of a Poisson process in which the intensity
function may be a random process (see Cox and Isham~\cite{cox1980point} for
an introduction).

We can formalize the description of the switching model  through the association
of standard Poisson
counting processes
with each potential state transition.  We thereby define, with associations,
the standard Poisson counting processes  $\Ydetb{i} (t) $  to represent detachment
of motor $ i$ and  $ \Yatt{i}
(t) $ to represent attachment of motor $ i $.  Then the dynamics of the attachment
states can be written, for $ i=1,\ldots,N $:
\begin{align}
d \Atstate{i}(t) &=\Yatt{i}
 \left(a^{(i)}  \int_0^t(1-\Atstate{i}(t^{\prime})) \, d t^{\prime}\right)
\\
 &-\Ydetb{i}
\left(\int_0^t\Atstate{i}(t^{\prime}) d_0^{(i)} \Upsilon^{(i)} \left(\kappa^{(i)}(X^{(i)}(t^{\prime})-Z(t^{\prime}))/F_d^{(i)}\right)\,
d t^{\prime}\right). \nonumber
\end{align}
Together with Eqs.~\eqref{motors}--\eqref{driftu} from Subsection~\ref{ssec:driftmodel},
we have a complete Markovian dynamical description for the motor-cargo model.

A common choice of $\Upsilon^{(i)}$, following the theory of Bell \cite{bell1978models},
is an  exponential function  (see \cite{muller2008tug,kunwar2010robust},
for instance). For simulations in Section~\ref{sec:sims}, we will  use
the more general double exponential detachment model, which is based on observations
that run length
is asymmetric
with respect to the direction of external load  \cite{milic2014kinesin,milic2017intraflagellar},
which can be argued to improve the processivity of a team of motors~\cite{kunwar2016anisotropy}.
Detachment rates for the double exponential detachment model  are given by
\begin{equation} \label{demodel}
d^{(i)}(F) = \begin{cases}d^{(i)}_{0-}\exp(-F/F_{d-}^{(i)}) & F \le
0,  \\
d^{(i)}_{0+}\exp(F/F_{d+}^{(i)}) & F >0.  \\

\end{cases}
\end{equation}
Here, $d^{(i)}_{0+}$ and $d^{(i)}_{0-}$ are, respectively, 
limits of detachment rates as the hindering (assisting) external force approaches
zero.  The corresponding force scales $ F^{(i)}_{d+} $ and $ F^{(i)}_{d-}$
of detachment are expressed in the literature~\cite{milic2014kinesin,milic2017intraflagellar}
in terms of characteristic length scales $\delta^{(i)}_+$ and $\delta^{(i)}_-$
via $ F_{d\pm}^{(i)} = k_BT/\delta^{(i)} $.  Note again our sign convention
on the force is opposite to how the experimental results are presented in~\cite{milic2014kinesin,milic2017intraflagellar},
but is consistent with the standard representation of force-velocity relations
in theoretical studies.
 In our one-dimensional model, we are not distinguishing between longitudinal
and transverse force components, which may have interesting dynamics from
geometric considerations of the cargo~\cite{Jamison:2010}.

From the additivity of Poisson rates,  a system with $N_u$ attached motors
has a constant total attachment rate of $\sum_{i:\Atstate{i} = 0} a^{(i)}$.
Under constant
detachment rates $d^{(i)},$ $i = 1, \dots, N$, which occur when $\Upsilon^{(i)}
$ is
a constant function,  switching between states is a homogeneous, continuous
time, finite state Markov chain, with an average time $\tau(\mathbf S)$ spent
in a state $\bAtstate = (\Atstate{1}, \dots, \Atstate{N})$ given by:
\begin{equation}
\tau(\mathbf S)= \frac{1}{\sum_{j=1}^N (1-\Atstate{j}) a^{(j)}+\sum_{j=1}^N
\Atstate{j} d^{(j)}}.
\end{equation}
 For more complicated detachment rates, the average time until either first
detachment or attachment admits no  explicit solutions. As we will see in
Section \ref{slowswitch},  however, under a slow switching regime, we  may
approximate detachment
rates by constants $ \bar d^{(i)}$ through averaging
over possible motor-cargo configurations.  This is in contrast to the quasi-steady
state model studied by Bressloff and Newby~\cite{bressloff2011quasi}, in
which transitions between
states are considered fast  compared to motor velocities.

\begin{table}
\centering
\caption{Typical values for kinesin-1 in water-like
  environments at saturating ATP concentrations. Values of parameters which
differ for kinesin-2 (specifically,
KIF3A/B) are in
  bold with parentheses~\cite{visscher1999single,furuta2013measuring,andreasson2015mechanochemical,mckinley2012asymptotic,milic2014kinesin,feng2018motor,milic2017intraflagellar}.}
\begin{tabular}{|l||l|l|}\hline \label{paramtable}
Parameter & Description&Typical values  \\\hline
$F_s^i$ & Motor stall force& 7 pN \cite{visscher1999single} \\\hline
$k_B T$ & Boltzmann constant by temperature& 4.1 pN nm\\\hline
$\kappa^{(i)}$ & Motor-cargo tether spring constant& 0.25 pN/nm
\cite{furuta2013measuring}
\\\hline
$v^{(i)}$  & Unencumbered motor velocity& 790 nm/s (\textbf{500 nm/s}) \cite{andreasson2015mechanochemical}\\\hline
$\gamma$ & Cargo friction& $1\times 10^{-5}$ pN s/nm \cite{mckinley2012asymptotic}\\\hline
$\gamma_{m,i}$ &  Motor friction& $1\times 10^{-6}$ pN s/nm\\\hline
$(\sigma^{(i)})^2 $ & Effective motor diffusion& 5000 nm\textsuperscript{2}/s
(\textbf{1500 nm\textsuperscript{2}/s})\cite{visscher1999single,andreasson2015mechanochemical}\\\hline
$F_T$ & Optical trapping force& -20 pN to 6 pN \cite{milic2014kinesin}
\\\hline
$a^{(i)}$ & Motor attachment rate & 4/s (\textbf{16/s}) \cite{feng2018motor}\\\hline
$d^{(i)}_{0-}$ & Small assisting force detachment rate  & 9.1/s \textbf{(5.6/s)
} \cite{milic2014kinesin,milic2017intraflagellar} 
\\\hline
$d^{(i)}_{0+}$ & Small hindering force detachment rate &  0.7/s \textbf{(2.3/s)}
\cite{milic2014kinesin,milic2017intraflagellar}\\\hline
$F^{(i)}_{d-} $& Assisting force detachment scale & 14 pN \textbf{(10 pN)}
\cite{milic2014kinesin,milic2017intraflagellar}\\\hline
$F^{(i)}_{d+} $& Hindering force detachment scale & 2.1 pN \textbf{(2.0 pN)}
\cite{milic2014kinesin,milic2017intraflagellar}\\\hline
\end{tabular}
\end{table}

\section{Nondimensionalization and averaging}
\label{sec:nond}
In this section, we prepare for the asymptotic analysis to coarse-grain the
detailed model from Section~\ref{sec:model} by a systematic nondimensionalization
in Subsection~\ref{sec:nondim}.  We thereby identify the detached motor and
cargo dynamics as faster than those of the attached motors, and conduct in
Subsection~\ref{sec:mavg} a stochastic averaging over the cargo and detached
motor coordinates to obtain an effective description only involving the attachment
state of the motors together with the spatial positions of attached motors.
 This section is essentially a generalization of the analysis from McKinley,
Athreya, et al~\cite{mckinley2012asymptotic}
to the case of nonidentical cooperative motors, together with a consideration
of the coarse-grained attachment or detachment events.

\subsection{Nondimensionalization}\label{sec:nondim}

For  the purposes of reducing the number of parameters in a motor-cargo system,
we perform a nondimensionalization of   (\ref{motors})-(\ref{driftu}), adapting
the nondimensionalizaton in~\cite{mckinley2012asymptotic}
for the case of identical cooperative motors.  To not only nondimensionalize
but normalize the variables for the asymptotic reduction exploiting time
scale disparities~\cite{lin1988mathematics,novozhilov2012fractional}, a reference
time scale of $ \gamma/\kappa $ and a reference length scale of $ \sqrt{2
k_B T/\kappa} $ was taken in the nondimensionalization, where $ \kappa $
was the common motor-cargo tether spring constant.  These characterize
the nominal fluctuation dynamics of the cargo, so the resulting nondimensional
equations are normalized to order unity for the cargo, and manifest the relatively
slow dynamics of the attached motors.  

To extend this nondimensionalization to nonidentical motors, we define
\begin{equation}
\bar \kappa = \sum_{j = 1}^N \kappa^{(j)}/N, \quad \tilde \kappa^{(i)} =
\kappa^{(i)}/\bar
\kappa,
\end{equation}
and take the average $ \bar \kappa $ to define the reference units in the
nondimensionalization.
Under the change of coordinates
$\tilde t = \bar \kappa t /\gamma$, and  
\begin{equation} \label{spacenondim}
\tilde X^{(i)}(\tilde t) = \frac{X^{(i)}(\gamma \tilde t/\bar \kappa)}{\sqrt{2k_BT/
\bar \kappa }}, \qquad \tilde Z(\tilde t) = \frac{Z(\gamma \tilde
t/\bar \kappa)}{\sqrt{2k_BT/
\bar \kappa }},
\end{equation}
equations (\ref{motors}) and (\ref{cargo}) may be written in nondimensional
form
\begin{align}
&d\tilde X^{(i)}(\tilde t) = \left( \epsilon^{(i)} g(s^{(i)}(\tilde X^{(i)}(\tilde
t)-\tilde
Z(\tilde t)))d\tilde t+\sqrt {\hat \rho^{(i)} \epsilon^{(i)}}dW^{(i)}(\tilde
t)\right)\Atstate{i}(\tilde t)\label{nondmotors}\\&+ \left ( -(\Gamma^{(i)})^{-1}\tilde
\kappa^{(i)}(\tilde
X^{(i)}(\tilde t)-\tilde Z(\tilde t))d\tilde t+(\Gamma^{(i)})^{-1/2}
\, dW^{(i)}(\tilde
t)\right)(1-\Atstate{i}(\tilde t)), \quad 1
\le i \le N,\nonumber
\\
&d\tilde Z(\tilde t)=\left( \sum_{j = 1}^N \tilde \kappa^{(j)}(\tilde X^{(j)}(\tilde
t)-\tilde Z(\tilde t))-\tilde{F}_T\right)d\tilde t + d W_z(\tilde
t) \label{nondcargo}.
\end{align}
The nondimensional attachment and detachment rates are now $\tilde a^{(i)}
=
a^{(i)}\gamma/\bar \kappa  $ and $\tilde d^{(i)} =d^{(i)}\gamma/\bar \kappa$.
Under this nondimensionalization, the detachment rate~\eqref{detach} then
becomes (expressed now as a stochastic process in time):
\begin{equation}
\tilde d^{(i)}(\tilde t) =  \tilde d^{(i)}_0\Upsilon (u^{(i)}(\tilde X^{(i)}(\tilde
t)-\tilde
Z(\tilde t)).  \label{nondetach}
\end{equation}
A listing of nondimensional parameters introduced in Eqs.~(\ref{nondmotors})-(\ref{nondetach})
and their typical values are provided in Table \ref{nondtable}. 

\begin{table}
\centering
\caption{Nondimensional groups and typical values for kinesin-1 and kinesin-2.
 When only a single value is given, it is common to both.  Otherwise the
kinesin-1 value is listed first, with the kinesin-2 value in bold text in
parentheses.  Attachment and detachment scales are taken for hindering forces.}

\begin{tabular}{|l|l|l|}\hline \label{nondtable}
Group & Definition &Typical value\\\hline
$\epsilon^{(i)} $&$\frac{ v ^{(i)}\gamma}{\sqrt{2k_BT\bar \kappa}}$ & $
6 
\times 10^{-3}  (\mathbf{3 
\times
10^{-3}})$ \\\hline
$s^{(i)} $&$\frac{\kappa^{(i)}}{F_s} \sqrt{\frac{2k_BT}{\bar \kappa}}$ &
$ 0.2$
\\\hline
$\tilde{F}_T $&$\frac{F_T }{\sqrt{2k_BT\bar \kappa}}$ &
-10 to 4 \\\hline
$\hat \rho^{(i)} $&$ \frac{(\sigma^{(i)})^2 \sqrt{\bar \kappa}}{v^{(i)}\sqrt{2k_BT}}$
 & $
1 $ (\textbf{.5}) \\\hline
$\Gamma^{(i)} $&$ \frac{\gamma_m^i}{\gamma}$  & $ 0.1$ \\\hline
$u^{(i)}$ &$\frac{\kappa^{(i)}}{F_d^i} \sqrt{\frac{2k_BT}{\bar \kappa}}$
&.7
 (\textbf{.7}) \\\hline
$\tilde a^{(i)}$ &$\frac{a^{(i)}\gamma}{\bar \kappa}$ &$2\times\ 10^{-4}$
($\mathbf{6\times\ 10^{-4}}$) \\\hline
$\tilde d^{(i)}$ &$\frac{d^{(i)}_0\gamma}{\bar \kappa}$ &$3\times\ 10^{-5}

(\mathbf{8\times\ 10^{-5}})$\\\hline
\end{tabular}
\end{table}

\subsection{Multiscale Averaging} \label{sec:mavg}
By comparing magnitudes of drift coefficients,  we are now able to identify
fast and slow variables.   For systems with multiple scales, a common method
of dimension reduction  averages out  fast variables by considering their
stationary distributions against fixed values of  slow variables (see Pavliotis
and Stuart~\cite{pavliotis2008multiscale}
for multiple examples).   We will use standard asymptotic
notation for indicating the relative magnitudes of quantities.  Informally,
$ f \ll g $ or equivalently $ f \sim o(g) $ means $ f $ is much smaller than
$ g $, $ f \sim g $ or equivalently $ f \sim \ord (g) $ means $ f $ and $
g $ are of comparable size, $ f \gg g $ means $ f $ is much greater than
$ g $, and $ f \sim O(g) $ means $ f $ is comparable to or smaller than $
g $.  All these concepts can be given precise asymptotic definitions~\cite{mhh:ipm},
which we follow though without making the formalism explicit.  From Table
 \ref{nondtable} and Eqs.~(\ref{nondmotors})
and (\ref{nondcargo}), we observe that a plausible asymptotic ordering for
the nondimensional parameters is:   $\tilde{a}^{(i)} \sim \tilde{d}^{(i)}
\ll \epsilon^{(i)}\ll1\ll
(\Gamma^{(i)})^{-1}$.   That
is, 
the dynamics for detached motors are faster  than  those for the cargo, which
in turn are faster than  those for attached motors, which in turn are faster
than the attachment and detachment processes.  We by no means claim this
asymptotic ordering is well satisfied for all molecular motors, or for kinesin-1
under all conditions, but simply that the assumptions on which our asymptotic
simplification is based is at least plausible based on the kinesin-1 data
 we have drawn from the literature, summarized in nondimensional form in
Table \ref{nondtable}.  The assumption that the attachment and detachment
processes are asymptotically slow compared to attached motor dynamics was
also adopted and exploited in Miles, Lawley, and Keener~\cite{miles2018analysis}.
 For mixtures of
motors, while we allow the attachment,
detachment, and unencumbered velocities to differ, we assume the parameters
in each group are of the same order of magnitude in our asymptotic ordering
($\tilde{a}^{(i)} \sim \tilde{a}^{(i^{\prime})} \sim \tilde{d}^{(i)} \sim
\tilde{d}^{(i^{\prime})}$,
$ \epsilon^{(i)} \sim \epsilon^{(i^{\prime})} $ for all $ 1 \leq i, i^{\prime}
\leq N$).

While the focus for this paper will mainly be for two motor systems, it is
possible  to write averaged formulas for both detached motor and cargo positions
under a generic system of $N$ motors.     Fixing a time $t\ge0$ and motor
index $i$, if  $\Atstate{i} (t)= 0$,  we may regard the distribution of 
the detached
motor $\tilde{X}^{(i)}(t)$ as approximately that of  the quasistationary
distribution
$p_{\tilde X^{(i)}|\tilde{Z}}$ 
under fast detached phase dynamics ($ \Atstate{i} (t) =0 $ in Eq.~\ref{nondmotors}),
with the slower cargo variable $\tilde Z(\tilde t)$ held at a fixed value
$ \tilde z$. This
is the
 Gaussian PDF
\begin{equation}\label{unboundstat}
 p_{\tilde X^{(i)}|\tilde{Z}}(\tilde x|\tilde z) = \sqrt {  \frac{\tilde
\kappa^{(i)}}\pi}\exp \left(-\tilde
\kappa^{(i)} (\tilde x -  \tilde z)^2\right),
\end{equation}
and all detached motor positions are conditionally independent given the
cargo
position $ \tilde Z (\tilde t) = \tilde z$.
Similarly, by fixing the positions of the slow attached motors, an approximation
of the distribution of the faster cargo position is the quasi-stationary
distribution
$p_{\tilde Z|\tilde{\mathbf{X}}^{(a)}}$ of (\ref{nondcargo}) with the states
of all motors
and the positions of the attached motors held at fixed values, which is another
Gaussian of the
form 
\begin{align}\label{zstat}
 &p_{\tilde Z|\tilde{\mathbf{X}}^{(a)},\bAtstate}(\tilde z |\mathbf
x,\batstate) \\
&= \sqrt{\frac{\sum_{j=1}^N \atstate{j}\tilde \kappa^{(j)}} {\pi}} \exp\left[
-\left(\sum_{j=1}^N \atstate{j}\tilde
\kappa^{(j)}\right) \left ( \tilde z- \left
[ \frac{\sum_{i=1}^N \atstate{j}\tilde \kappa^{(j)}\tilde x^{(j)} - \tilde
F_T}{\sum_{i=1}^N \atstate{j}\tilde
\kappa^{(j)} }
\right]\right)^2\right], \nonumber
\end{align}
where $ \mathbf x = (x^{(1)}, x^{(2)},\ldots,x^{(N)}) $ and $ \batstate =
(\atstate{1},\atstate{2},\ldots,\atstate{N})
$ parameterize, respectively the positions and states of all $N$ motors.
Note that the above formula for the quasi-stationary distribution of
the cargo does not actually depend on the positions of the detached motors
(indices $i $ for which $ \atstate{i} =0$), as these are fast relative to
the cargo,
and so are treated as already averaged out on the cargo time scale.  This
is why we denote the fixed variable as $ \tilde{\mathbf{X}}^{(a)} $, though
we write the quasi-stationary distribution formally as a function of all
motor
positions to keep notation simple.  Moreover, the averaging of the detached
motors does
not affect the cargo dynamics to leading order, because the tether force
will average to zero, and the contribution of the force fluctuations to the
cargo diffusivity are 
$ O(\Gamma^{(i)}) $.  The average position for the
cargo according to Eq.~\eqref{zstat} is
a weighted average of the attached motor positions shifted by a multiple
(which would be the simple inverse of the number of attached motors if they
had the same tether spring constants) of the nondimensional trap
force.   In reality the cargo should lag a bit from this weighted average
position of the attached motors because of balancing the viscous drag force.
 Our treatment of $ \epsilon^{(i)} $ as a small parameter, however, implies
that the cargo drag force is being treated as small compared to the force
scale of the thermal fluctuations of the cargo, so this mean lag would be
small compared to the standard deviation of the cargo fluctuations and is
thus neglected.
We will often refer
to (\ref{zstat}) in the specific case of systems with $N=2 $ motors.  With
 one out of the two motors attached,  
\begin{align}\label{zstat1}
 p_{\tilde Z|\tilde{\mathbf{X}}^{(a)},\bAtstate}(\tilde z|(\tilde x^{(1)},\tilde
x^{(2)}),(1,0)) &=
\sqrt{\frac{ \tilde
\kappa^{(1)}} {\pi}} \exp\left[ - \tilde
\kappa^{(1)} \left ( \tilde z-  \tilde x^{(1)}
+ \frac{\tilde F_T}{ \tilde
\kappa^{(1)}}\right)^2\right], \\
p_{\tilde Z|\tilde{\mathbf{X}}^{(a)},\bAtstate} (\tilde z|(\tilde x^{(1)},\tilde
x^{(2)}),(0,1)) &=
\sqrt{\frac{ \tilde
\kappa^{(2)}} {\pi}} \exp\left[ - \tilde
\kappa^{(2)} \left ( \tilde z-  \tilde x^{(2)}
+ \frac{\tilde F_T}{ \tilde
\kappa^{(2)}}\right)^2\right],  \nonumber
\end{align}
and with both motors attached,
\begin{equation}\label{zstat2}
p_{\tilde Z|\tilde{\mathbf{X}}^{(a)},\bAtstate}(\tilde z|(\tilde x^{(1)},\tilde
x^{(2)}),(1,1)) = \sqrt{\frac{2}
{\pi}} \exp \left [-2   \left
(\tilde z- \frac{\tilde \kappa^{(1)} \tilde x^{(1)} + \tilde \kappa^{(2)}
\tilde x^{(2)}}{2}
+ \frac{\tilde F_T}{2}\right)^2\right].
\end{equation}

Our nondimensionalization set the time scale to be order unity for the cargo,
so the dynamics of the slower attached motors expressed in Eq.~\eqref{nondmotors}
appears weak on this time scale
($\ord (\varepsilon^{(i)}) $ drift and diffusivity for attached motors $i$,
with $ \varepsilon^{(i)} \ll 1 $). 
Thus, to see nontrivial attached motor dynamics, we must go to a longer time
scale 
\begin{equation}
\tilde t = \bar t/\bar \epsilon, \quad \bar \epsilon =
\sum_{j=1}^N \epsilon^{(j)}/N, \label{eq:tbar}
 \end{equation}
 over which the cargo now appears $ \ord((\bar \epsilon)^{-1})$
fast
and equilibrates quickly relative to changes in attached motor position.
  From stochastic averaging theory (see~\cite{skorokhod2002random,liptser1990stochastic}),
we can approximate the attached motor positions on this time scale $\tilde
X^{(i)}(\bar t/\bar \epsilon)$ by an averaged stochastic process $\bar
X^{(i)}(\bar t)$ satisfying the system of SDEs
  \begin{equation}\label{xbar}
d\bar X^{(i)}(\bar t) = \bar g^{(i)}(\{\bar X^{(i)}\}_{i = 1}^N(\bar t);\{\Atstate{i}\}_{i
=
1}^N) d\bar t +\sqrt
{\rho^{(i)}} d W^{(i)}(\bar t), \quad \rho^{(i)} = \frac{\epsilon^{(i)}}{\bar
\epsilon}\hat
\rho^{(i)},
\end{equation}
with an effective drift obtained by averaging over the  quasi-stationary
distribution of the cargo:
\begin{equation}\label{avg}
\bar g^{(i)}(\mathbf x; \mathbf s) = \frac{\epsilon^{(i)}}{\bar \epsilon}\int_{\mathbb{R}}
g(s^{(i)}(\tilde x^{(i)}-\tilde z))
p_{\tilde Z|\tilde{\mathbf{X}}^{(a)},\bAtstate} (\tilde z|\mathbf x, \batstate)d\tilde
z.
\end{equation}
Note the averaged dynamics of the attached motors in Eq.~\eqref{xbar} are
coupled directly to each other, through the averaging out of the cargo variable
to which they are explicitly coupled in~\eqref{nondmotors}.  The averaged
drift $ \bar g^{(i)} $ ostensibly depends on all motor positions, but it
actually
is independent of the detached motor positions because the same is true of
$p_{\tilde Z|\tilde{\mathbf{X}}^{(a)},\bAtstate}$ in Eq.~\eqref{zstat}. 
Bouzat \cite{bouzat2016models} proposes alternatively to coarse-grain the
effects of the cargo fluctuations on the effective velocity of motors through
an exponential time-averaging of the force felt from the cargo.  This should
give equivalent results to the more straightforward stochastic averaging
used here for the ``robust'' regime of averaging time scales advocated in~\cite{bouzat2016models}.

Continuing with our assumption that the detachment process is slow compared
to the time scale of motor motion  ($\tilde d^{(i)}\ll \epsilon^{(i)}$),
we may also
obtain an  averaged detachment rate $\bar d^{(i)}(\mathbf{x})$ for each motor
$i$ (within the framework~\cite{tk:ampsa} of averaging for general
Markov processes): 
\begin{align}\label{dbarxz}
 \bar{d}^{(i)} (\mathbf x,\mathbf s)= 
 \frac{\tilde d_0^{(i)}}{\bar \epsilon} \int_\mathbb R \Upsilon\left(u^{(i)}\left
(\tilde x^{(i)}-\tilde z
\right)\right)p_{\tilde Z|\tilde{\mathbf{X}}^{(a)},\bAtstate} (\tilde z|\mathbf
x, \batstate) d\tilde z. 
\end{align}
This formula may be viewed as a more detailed implementation of the idea
given in~\cite{lipowsky2009myosin,arpag2019kin1kin3} that the effective detachment
rate for
a motor should be computed by averaging the nominal force-detachment rate
formula over a probability distribution of forces felt from the cargo, and
that this can lead to a significant enhancement relative to the evaluation
of the detachment rate in terms of simply the average force felt from the
cargo.  The constant attachment rates on the longer time scale~\eqref{eq:tbar}
are
now 
\begin{equation}
 \bar{a}^{(i)}=  \tilde{a}^{(i)}/\bar \epsilon.
 \end{equation}

When a detached motor attaches, its attachment position is taken to be drawn
from the quasi-stationary distribution of the detached motors, given
fixed
positions of the attached motors.  This follows from our modeling of the
attachment process as independent of spatial configuration in the present
work. 
\cite{furuta2013measuring,bouzat2016models,kunwar2016anisotropy} place attaching
motors uniformly
at random over the region of the microtubule where the motor-cargo tether
is unstretched; this would be roughly consistent with our approach when the
linear spring is modified to give no resistance to compression below a rest
length (see Appendix~\ref{app:nonlinspring}).
\cite{vershinin2018intersect}~places attaching motors at the closest position
on the microtubule to their attachment point to the cargo (which, with our
point particle representation of the cargo, would here amount to attaching
the motor at the current cargo position $ \tilde Z $).  
 Because
of the assumed time scale separation, the relative position of the detached
motors and the cargo is independent of the relative position of the cargo
and the attached motors, and these are governed by the respective Gaussian
quasi-stationary distributions~\eqref{unboundstat} and~\eqref{zstat1}.  The
quasi-stationary distribution of the position of the detached motors for
fixed positions of the attached motors is therefore obtained by convolving
these Gaussian quasi-stationary distributions, giving
rise to a Gaussian distribution as well.  
In the simple case of $N=2$ motors with exactly one of the two motors
currently attached, the probability densities for the attachment position
of the detached motor would be:
\begin{align}\label{patt}
 p^{(a)}_{\tilde{X}^{(2)}} (x^{\prime}|(\tilde x^{(1)},\tilde x^{(2)}), (1,0))
 &= \int_{\mathbb
R} p_{\tilde X^{(i)}|\tilde Z}(x^{\prime}|\tilde z)p_{\tilde Z|\tilde{\mathbf{X}}^{(a)},\bAtstate}
(\tilde z|(\tilde x^{(1)}, \tilde x^{(2)}), (1,0)) d\tilde z\\&= \sqrt {\frac{\tilde
\kappa^{(1)} \tilde \kappa^{(2)}}{2\pi}}\exp
\left(-\frac{\tilde \kappa^{(1)} \tilde \kappa^{(2)}}{2}\left(x^{\prime}
-\tilde
x^{(1)}+  \frac{\tilde
F}{\tilde \kappa^{(1)}}\right)^2\right), \\
p^{(a)}_{\tilde{X}^{(1)}} (x^{\prime}|(\tilde x^{(1)},\tilde x^{(2)}), (0,1))
 &= \sqrt {\frac{\tilde \kappa^{(1)} \tilde \kappa^{(2)}}{2\pi}}\exp
\left(-\frac{\tilde \kappa^{(1)} \tilde \kappa^{(2)}}{2}\left(x^{\prime}
-\tilde
x^{(2)}+  \frac{\tilde
F}{\tilde \kappa^{(2)}}\right)^2\right). 
\end{align} 
For the general case of $ N$ motors, the attachment position of a currently
detached motor $ i $ would have a Gaussian probability distribution with
mean
\begin{equation*}
\frac{\sum_{j= 1}^N \atstate{j}\tilde \kappa^{(j)}\tilde x^{(j)}-\tilde F_T}{\sum_{j=
1}^N\atstate{j} \tilde \kappa^{(j)}}
\end{equation*}
and variance
\begin{equation*}
\frac{1}{2\sum_{j= 1}^N
\atstate{j}\tilde \kappa^{(j)} }+\frac 1{2\tilde \kappa^{(i)}}.
\end{equation*}


\section{Effective dynamics under slow switching approximation}\label{slowswitch}

\begin{figure}
\centering
\includegraphics[width=\textwidth]{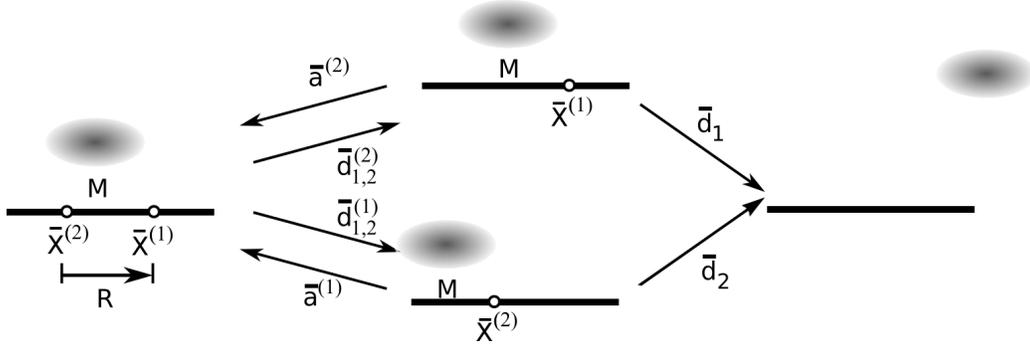}
\caption{\textbf{Switched diffusion model. } 
Effective dynamics of cargo with two attached motors on time scale  $\tilde{a}^{(i)}
\sim \tilde{d}^{(i)}
\ll \epsilon^{(i)}\ll \tilde{t} \ll 1\ll
(\Gamma^{(i)})^{-1}$ long compared to detached motor and cargo fluctuation
dynamics but short compared to attached motor dynamics.  The cargo dynamics
are represented on this time scale by a cargo tracking variable $M $, as
discussed in Subsection~\ref{ssec:track}.
Left: A cargo with both motors attached to the microtubule.  The cargo tracking
variable dynamics~\eqref{twomotorsdem} and the detachment rate  $\dettwoz{i}$~\eqref{deffzmotori}
of each motor $i $
depends on the displacement $R $ between the attached motors, also evolving
dynamically.  Middle: With only one motor attached, 
the cargo tracking variable $  M $ with one attached motor evolves with constant
drift~\eqref{gbari} and diffusivity~\eqref{dbari}.
 From these states, the system either detaches completely (as shown in the
rightmost part of the figure), with effective rate given in~\eqref{singledbar}
or
returns to having two attached motors, with rates $\bar{a}^{(i)}$~\eqref{singleabar}.
}.\label{switch2}
\end{figure}

\begin{figure}
\centering
\includegraphics[width=1.1\textwidth]{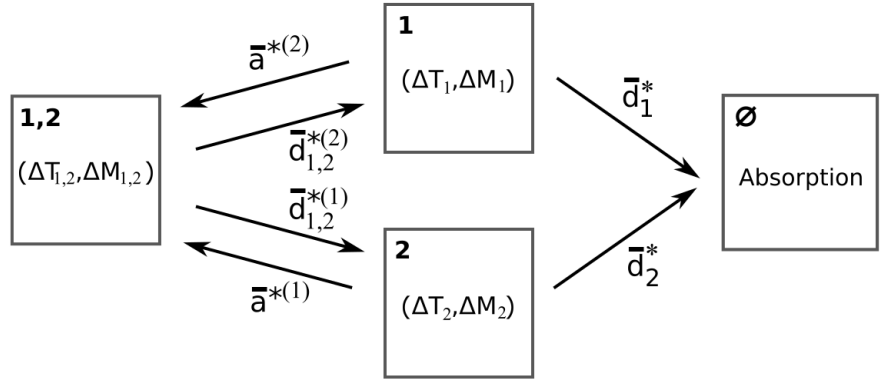}
\caption{\textbf{Coarse-grained Markov chain model.} Under the assumption
of slow switching, the switched diffusion model (see Fig. \ref{switch2})
is further coarsened by averaging intermotor separation.  The random duration
$ \Delta \Tstate{\attmots} $ spent in state $ \attmots $ is exponentially
distributed with mean determined in the usual way by the transition rates
out the state.  The random displacement
$ \dMstate{\attmots} $ for each attachment state $\attmots $ (shown in the
upper left corner of each box) is then found
through (\ref{mchange}).  Detachment
rates $\dettwo{i}$ from the state of two attached motors, given by (\ref{deffmotori}),
are now constants coarse-grained with respect to intermotor separation.}\label{switch3}
\end{figure}

From here on, we will focus on dynamics for a two motor ($N=2$) system during
a processive
run. At any one time, either one or two motors may be attached to the microtubule,
which gives three possible states, or the run may terminate when both motors
are simultaneously detached from the microtubule.  The dynamical description
for $N=2 $ motors resulting from the coarse-graining over the dynamics of
detached motors and cargo fluctuations is summarized in Figure~\ref{switch2}.
 The cargo position is now represented as a probability distribution given
the location and identity of the attached motors (see Eq.~\eqref{mdef}),
with a ``cargo
tracking variable'' $ M(\bar{t}) $ representing the conditional mean position
of the cargo.  The symmetry of the dynamics under common spatial translation
of all entities implies the effective dynamics while in the state with one
motor attached has  constant drift and attachment/detachment rates, while
these quantities in the state with two motors attached only depends on the
directed separation of the attached motors 
 \begin{equation}
 R(\bar t) \equiv \bar X^{(1)}(\bar t)-\bar X^{(2)}(\bar t). \label{eq:Rdef}
 \end{equation}
 
In this section,  we will make use of the assumption
that the attachment and detachment rates are slow relative to the attached
motor dynamics ($\tilde a, \tilde d \ll \epsilon $), to homogenize the spatial
dynamics within each attachment state and thereby further coarse-grain our
model, based on stochastic differential equations with configuration-dependent
detachment rates, into simple continuous-time
Markov chain dynamics on the state space of attachment states, together with
the cargo tracking variable $ M(t) $ which may be thought of as an accumulated
reward
function associated with the Markov chain. The four states of this  Markov
chain are labeled
by a list $ \attmots $ of the indices of the attached motors,  as indicated
in Figure \ref{switch3}.  
 The transition rates between states after homogenization of attached motor
dynamics are now all constant, also
indicated
in Figure \ref{switch3}.

 The state $ \attmots = \emptyset $ acts as an absorbing state terminating
the processive run.  In each the other three states, the cargo tracking variable
undergoes a constant-coefficient drift-diffusion dynamics.
  Thus, a visit to a state
$ \attmots \neq \emptyset $ is associated to a cargo tracking variable increment
\begin{equation*}
 \dMstate{\attmots} 
 = \Vstate{\attmots} \Delta\Tstate{\attmots} + \sqrt{2
\Dstate{\attmots}}
\Delta W (\Delta\Tstate{\attmots}),
 \end{equation*}
 where $ \Delta \Tstate{\attmots} $ is the duration of the visit to state
$ \attmots
$, $ \Vstate{\attmots}$ and $ \Dstate{\attmots} $ are, respectively, the
constant coarse-grained velocity and diffusivity in state $ \attmots $,
and $ \Delta W (t)\sim N(0,t)$ denotes the increment of the Wiener process
over a time $ t $.  In other words, the cargo tracking variable increment
is Gaussian conditioned on the duration $ \Delta \Tstate{\attmots} $ in the
state $ \attmots$:
\begin{equation}
 \dMstate{\attmots}|\dTstate{\attmots} \sim N(\Vstate{\attmots} \Delta\Tstate{\attmots},2
\Dstate{\attmots} \dTstate{\attmots}). \label{mchange}
 \end{equation}
Here and later, the notation $ Y\sim N(\mu,\sigma^2)$ indicates that $ Y
$ is a normally distributed random variable with
mean $ \mu $ and variance $ \sigma^2 $.  The tracking variable $ M $ will
also suffer jumps associated to transitions, corresponding to adjustments
in the mean cargo position under the new configuration of attached motors.

We now proceed to more precisely quantify the elements of the coarse-grained
Markov chain description depicted in Figure~\ref{switch3}.
The cargo tracking variable $ M $ for the motor-cargo complex in the coarse-grained
representation is defined and discussed in Subsection~\ref{ssec:track}. 
Thereafter, formulas
for the effective drift and diffusion coefficients of the cargo tracking
variable in each state ( $ \Vstate{\attmots} $ and $ \Dstate{\attmots} $)
are presented in Subsection~\ref{ssec:homtwomotor}.
In Subsection \ref{sec:effswitch}, we describe both the transition rates
of the coarse-grained Markov chain as well as the associated jumps in the
tracking variable at a transition.  These coarse-grained quantities are expressed
following the nondimensionalization from Subsection~\ref{sec:nondim}, followed
by the passage to the long time scale $ \bar t = \tilde t \bar \epsilon $,
so the effective drift, diffusion, and attachment/detachment rate formulas
presented would need to be multiplied by $  \bar{\epsilon} $ to give their
expressions in terms of the original nondimensionalization.
Note that when describing effective transport and switching rates
from various states of attachment, we will typically list the indices of
the attached motors in the subscript (without parentheses or brackets),
and when needed,
indicating by a parenthesized superscript the index of which motor
in that state is being
described by the parameter.  Finally, in Subsection~\ref{sect:procdisp},
we
shift focus to a more event-based view, decomposing the progress of a motor-cargo
complex along a microtubule in terms of cycles of detachment and reattachment
of a motor before eventual complete detachment of both motors.  This representation
of the motor-cargo complex dynamics in terms of these cycles will be the
basis for computing the overall statistics of cargo transport in Section~\ref{sec:renewal}.

\subsection{Tracking Variable for Coarse-Grained Motor-Cargo Complex} \label{ssec:track}

One challenge in characterizing the progress of the motor-cargo system through
phases of attachment and detachment is the choice of a ``tracking'' variable
that remains well-defined in the various states of motor attachment.  The
cargo variable $ \tilde Z (\tilde t) $ is an obvious candidate since in experiments
it is the largest and most easily observed object, and whose progress through
space is of practical importance.  Mathematically, though, as shown in the
dimensional analysis from Subsection~\ref{sec:nondim}, the cargo will tend
to fluctuate more rapidly than attached motors, which makes it somewhat awkward
to use its instantaneous position as a tracking variable on long time scales.
We therefore introduce  the variable $\tilde M (\tilde t)$ for the mean
cargo position at time $\tilde t$ under the quasi-stationary distribution
$p_{\tilde
Z|\tilde{\mathbf{X}}^{(a)},\bAtstate}$~\eqref{zstat} for the cargo given
the current positions of the
motors attached to the microtubule at time $ \tilde t $:
\begin{align}
\tilde M (\tilde t) &\equiv 
\int_{-\infty}^{\infty} \tilde z p_{\tilde Z|\tilde{\mathbf{X}}^{(a)},\textbf
S}(\tilde z |\tilde{\mathbf{X}} (\tilde t),\bAtstate (\tilde t)) \nonumber
\\&= \left
[ \frac{\sum_{j = 1}^N \Atstate{j} (\tilde t)\tilde \kappa^{(j)}\tilde X^{(j)}
(\tilde
t)}{\sum_{j = 1}^N\Atstate{j} (\tilde t)\tilde
\kappa^{(j)}}
- \frac{\tilde F_T}{\sum_{j = 1}^N \Atstate{j} (\tilde t)\tilde
\kappa^{(j)}}\right].
 \label{mdef}
\end{align}
One can check that $ \tilde M (\tilde t) $ is equivalent to the deterministic
mechanical equilibrium of the cargo, for given attached motor positions,
in the absence of stochastic fluctuations.  Markov chain models for cargo
transport with attaching and detaching motors often model the cargo as always
being exactly at this  position $ \tilde M (\tilde t)$ of force balance relative
to the attached motors~\cite{klumpp2005cooperative,klumpp2015review}; our
present model accounts for cargo
fluctuations about this mechanical equilibrium.  
Nonetheless, $\tilde M$ is a more convenient variable for tracking the progress
of the cargo through episodes of attachment and detachment. Moreover, the
cargo position $ \tilde Z (\tilde t)$, under our time scale separation assumptions,
has a Gaussian distribution centered at $ \tilde M (\tilde t) $ with standard
deviation no larger than $ \max_{1 \leq j\leq N} \{\frac{1}{2\tilde{\kappa}^{(j)}}\}
$ (see Eq.~\eqref{zstat}).  Consequently, the long-time statistical transport
properties of the cargo position $ \tilde Z (\tilde t) $ and the tracking
variable $ \tilde M (\tilde t) $ are equivalent.  Note below we define the
tracking variable on the longer time scale (which is being used after averaging
out the cargo dynamics in Subsection~\ref{sec:mavg}) as
\begin{equation}
M (\bar t) = \tilde M (\bar{t}/\bar{\epsilon}). \label{eq:trackresc}
\end{equation}

\subsection{Effective Drift and Diffusion Coefficients in Attachment States}\label{ssec:homtwomotor}
For states where the motor with index $ i $ is attached but the other motor
is detached, the effective drift is:
\begin{equation}\label{gbari}
\Vstate{i}= \sqrt {\frac{\tilde \kappa^{(i)}}\pi}\frac{\epsilon^{(i)}}{\bar
\epsilon}
\int_\mathbb R g(s^{(i)} y)\exp(-\tilde \kappa^{(i)}(y-\tilde F_T/\tilde
\kappa^{(i)})^2)dy.
\end{equation}
and the effective diffusivity is
\begin{equation}
\Dstate{i}=\frac{1}{2} \rho^{(i)}. \label{dbari}
\end{equation}
These formulas follow directly from Eq.~\eqref{xbar},  once we recognize
from Eq.~\eqref{mdef}, that
 \begin{equation}
 \tilde M(\tilde  t)  =\tilde X^{(i)}(\tilde t)-\tilde
F/\tilde \kappa^{(i)}. \label{trackvar1}
\end{equation}
Now we just carry over the result from Eq.~\eqref{xbar}, noting the expression~\eqref{avg}
is constant for the case of one attached motor.

For the state where both motors are attached, the effective drift is: 
\begin{equation} \label{eq:barvtwo}
\Vstate{1,2} = \int_\mathbb R  G_+(r) p_R(r)dr.
\end{equation}
where
\begin{equation}\label{statr}
p_R(r) = C_R \exp \left [\frac{2}{ \rho^{(1)} +\rho^{(2)}}\int_0^r G_-(r')dr'\right
], \quad -\infty<r<\infty,
\end{equation}
with normalizing constant $ C_R $, is the stationary probability density
for the displacement $ R = X^{(1)}-X^{(2)} $ between the two attached motors.
 The effective diffusivity of the motor-cargo complex in this state is:
\begin{align}
& \Dstate{1,2} = \int_{-\infty}^\infty \left(\frac{2}{\rho^{(1)}+\rho^{(2)}}\right)\left(\int_{-\infty}^r
( G_+(r^{\prime}) -\barvtwo)p_R(r^{\prime})d r^{\prime}\right)^2\frac{1}{p_R(r)}\,
dr \nonumber\\&+ 
\left(\frac{\rho^{(1)}\tilde\kappa^{(1)}-\rho^{(2)}\tilde\kappa^{(2)}}{\rho^{(1)}+\rho^{(2)}}\right)\int_\mathbb
R ( G_+(r) -\barvtwo)p_R(r)\cdot rdr+\frac{\rho^{(1)}(\tilde
\kappa^{(1)})^2+\rho^{(2)}(\tilde\kappa^{(2)})^2}8. \nonumber
\end{align}
The auxiliary functions referenced in these formulas are:
\begin{align}
 \quad G_+(r) &=\frac{\tilde \kappa^{(1)}}{2} G^{(1)}(r) +\frac{\tilde
\kappa^{(2)}}{2}G^{(2)}(r),\\
G_-(r) &= G^{(1)}(r) - G^{(2)}(r).
\end{align}

The derivation of these formulas for the effective transport for the state
with two motors attached proceeds as follows.  For known motor positions
$x^{(1)},
x^{(2)}$,  we can express the cargo-averaged drift coefficients purely in
terms
of the signed displacement $ r = x^{(1)} - x^{(2)} $ between the motors:
\begin{align}\label{giform}
&\bar g^{(1)}(x^{(1)},x^{(2)};(1,1)) = G^{(1)} (x^{(1)}-x^{(2)}), \\
&\bar g^{(2)}(x^{(1)},x^{(2)};(1,1)) = G^{(2)} (x^{(1)}-x^{(2)}), \\
&G^{(i)}( r)= \frac{\epsilon^{(i)}}{\bar \epsilon} \sqrt{\frac
2 \pi}\int_{\mathbb{R}} g(s^{(i)} y) \exp\left( -2\left (
y+  \frac{ (-1)^{i}\tilde \kappa^{(i')}r}2-\frac{\tilde F_T}2 \right)^2\right)dy.
\end{align}
 This representation is achieved by the change of variable $y =x^{(i)}-z$
in (\ref{avg}).\\

The tracking variable~\eqref{mdef} in this state is, from Eq.~\eqref{zstat2},
\begin{equation}
\tilde M(\tilde t)  =
(\tilde \kappa^{(1)}\tilde X^{(1)}(\tilde t)+\tilde \kappa^{(2)}\tilde X^{(2)}(\tilde
t)-\tilde F_T)/2, \label{trackvar2}
\end{equation}
Passing now to the long time scale $ \bar t/\epsilon $, we can recast the
cargo-averaged dynamics~\eqref{xbar} for the two attached motors in terms
of
this tracking variable rescaled to large time, $ M(\bar t) $~\eqref{eq:trackresc},
 and the (signed) intermotor separation
 \begin{equation*}
 R (\bar t) \equiv \tilde X^{(1)} (\bar t/\bar \epsilon) - \tilde X^{(2)}
(\bar t/\bar \epsilon)
 \end{equation*}
which gives  
\begin{align}
d M(\bar t) &= G_+( R(\bar t)) d \bar t+\frac {\sqrt{\rho^{(1)}
(\tilde\kappa^{(1)})^2}}2dW^{(1)}(\bar t)+\frac {\sqrt{\rho^{(2)}(\tilde\kappa^{(2)})^2}}2
dW^{(2)}(\bar t), \label{twomotorsdem}\\ 
d R(\bar t)&= G_- (R(\bar t))dt+\sqrt{\rho^{(1)}}dW^{(1)}(\bar t)-\sqrt{\rho^{(2)}
} dW^{(2)}(\bar
t). \label{twomotorsder}
\end{align}
Under  the regime where switching is slow relative to detachment,
we may 
  simply homogenize the internal variable $R$ in order to obtain the effective
velocity and diffusion for $M$ on the long time scales when attachment or
detachment occurs. As (\ref{twomotorsder}) does not depend on
$M$, the stationary distribution for the process $R$ is a potential function,
given in Eq.~\eqref{statr}.
The formula~\eqref{eq:barvtwo} follows directly.

The computation for effective diffusivity is more complicated.  For nonidentical
motors, the driving terms in the stochastic differential equations in Eqs.~(\ref{twomotorsdem})-(\ref{twomotorsder})
are correlated in general, unlike the identical motor case from~\cite{mckinley2012asymptotic}.
 
Only minor
modifications are needed to generalize the derivation for effective diffusion
 found in Pavliotis and Stuart~\cite{pavliotis2008multiscale} for the case
of uncorrelated stochastic
driving, as we show in Appendix~\ref{app:twomotorapp}.

 \subsection{Effective switching dynamics} \label{sec:effswitch}
After presenting in Subsection~\ref{subsub:effdet} the effective switching
rates between attachment states (as indicated in Figure~\ref{switch3}), we
discuss the effectively instantaneous jumps in the cargo tracking variable
that occurs in the coarse-grained representation when switches occur.
Namely, when a motor $i $ detaches from the two-motor-attached state (transition
$(1,2) \rightarrow (i^{\prime})$), the tracking variable undergoes a jump
$ \dMdetjump{i} $ given by equation (\ref{jumpoff}).  When a motor $ i $
attaches to form the two-motor-attached
state (transition $ (i^{\prime})\rightarrow (1,2) $), the tracking variable
undergoes a jump $ \dMattjump{i}$, given by equation (\ref{jumpon}).

\subsubsection{Effective detachment rates} \label{subsub:effdet}
From our assumptions about the detachment time scale being slower than the
attached motor time scale ($\tilde{d}^{(i)} \ll \epsilon^{(i)} $)  in Subsection~\ref{sec:mavg},
we can apply stochastic averaging~\cite{tk:ampsa} to  approximate  the rate
at which motor $ i $ detaches from a state with both motors attached 
by  averaging over the stationary distribution $p_R (r) $~\eqref{statr} for
the motor
separation $ R $, just as we did for the effective velocity:
\begin{equation}\label{deffmotori}
\dettwo{i}=\int_\mathbb{R} \dettwoz{i}(r) 
 p_R(r)dr,
\end{equation}
\begin{align}\label{deffzmotori}
\dettwoz{i}(r) =\sqrt{\frac2
\pi}\frac {\tilde d^{(i)}_0}{\bar \epsilon} \int_\mathbb R \Upsilon\left(u^{(i)}y\right)
\exp\left(-2\left(y- \frac
12((-1)^{i+1} \tilde \kappa^{(i')}r -\tilde F_T)\right)^2\right)dy.
\end{align}
This last formula is just an expression of the detachment rate (\ref{dbarxz})
solely through the intermotor distance $r = x^{(1)}-x^{(2)}$,
An effective total rate of detachment from the state of both motors attached
is then  
\begin{equation}\label{deff}
\dettwoboth = \dettwo{1}+\dettwo{2}.
\end{equation}

 In the case of a single attached motor with index $i$, we have
a constant attachment
rate $\tilde{a}^{(i')}$, and detachment rate   $\tilde d^{(i)}( x^{(i)},z)$
that
is dependent
on the attached motor position $x^{(i)}$ and cargo position $z$.  On the
longer
$ \bar{t} = \bar{\epsilon} \tilde{t} $ time scale, the slow switching approximation
would reduce the detachment rate~\eqref{dbarxz} to a constant:
\begin{equation}\label{singledbar}
  \detone{i}=\sqrt{\frac{\tilde
\kappa^{(i)}}{\pi}}\frac{\tilde d^{(i)}_0}{\bar \epsilon} \int_{\mathbb
R}\Upsilon
(u^{(i)}y)\exp(-\tilde\kappa^{(i)}(y- \tilde F_T /\tilde \kappa^{(i)})^2)dy.
\end{equation}
and simply rescale the constant attachment rates:
  \begin{equation}
    \bar{a}^{*(i)} \equiv \frac{\tilde{a}^{(i)}}{\bar \epsilon}.\label{singleabar}
    \end{equation}

\subsubsection{Jumps at detachment from state of both motors attached} \label{subsub:detach}
When a motor detaches from a state of two attached motors, there is an immediate
change
in the force balance between motors and cargo.    The cargo, which is fast
relative to a single bound motor, quickly adjusts to the new motor configuration.
  This  results in a jump  of the tracking variable position. A similar readjustment
occurs with motor attachment. See Fig. \ref{fig:simjump} in Section~\ref{sec:sims}
for simulations
depicting jumps in cargo positions.
In this subsection, we describe distributions of jump sizes at these switching
events. This is done under the assumption of slow switching, so that we may
assume the intermotor distance variable $ R (\bar t) $ in the state of two
attached
motors (in addition to the cargo) has achieved its stationary distribution.
 
 We focus in this subsection on the statistical behavior of the system at
a  random time $ \dettime $ at which one of the two motors detaches ($\bAtstate
(\dettime^{-}) = (1,1) \neq \bAtstate (\dettime) $).  
The distribution of $ R (\dettime^-) $ just before detachment will not be
the same as the stationary
distribution of $ R(t) $ due to the dependence of the detachment rate on
$ R(t) $.  Rather, the distribution of the intermotor distance just before
first detachment
$ \Rdet = R(\dettime^-)  $ will be reweighted by the detachment rate, yielding
the probability
density:
\begin{align}\label{rdetach}
p_{\Rdet} (r) 
= \frac{(\dettwoz{1}(r)+\dettwoz{2}(r))p_R(r)}{\dettwoboth}.
\end{align} 
This can be readily argued by considering a short time interval of length
$ \Delta t $ over which fluctuations in $ R(t) $ are negligible, using Bayes'
rule to calculate the conditional probability of $ R(t) $ given that detachment
occurs during the time interval, and passing to the limit $ \Delta t \downarrow
0 $.



Next, we denote by $ \detmot $ the index of the motor which first detaches
from the two-motor-attached state (at the random time $ \dettime $).  From
the standard theory of continuous-time jump processes,
\begin{equation}
 \mathbb{P} (\detmot=i|R(\dettime)=r) = \frac{\dettwoz{i}(r)
}{\dettwoz{1}(r)+\dettwoz{2}(r)}, \label{conddet}
\end{equation}
and thus the unconditional probability that motor $i $ detaches first is:
\begin{equation} \label{probdetone}
\pdet{i} \equiv \mathbb P(\detmot=i) = \int_{-\infty}^\infty  \frac{\dettwoz{i}(r)
}{\dettwoz{1}(r)+\dettwoz{2}(r)} p_{\Rdet}(r)dr =  \frac{\dettwo{i}
}{\dettwoboth}.
\end{equation}
This is consistent with the coarse-grained description of the attachment
states of the motors, under the slow switching approximation, having
the properties of a continuous-time Markov chain.

%
 
 The jump of the tracking variable $M$ at detachment  is essentially a result
of how the mean position of cargo relies upon the number of attached
motors.  This is represented by  the difference between equations (\ref{trackvar1})
and (\ref{trackvar2}). Consider, for now, that motor 2 detaches ($\detmot=2$)
from the two-motor-attached state at time $ \dettime $.
The jump size $ \dMdetjump{2} = M(\dettime) - M(\dettime^{-}) $ will be
\begin{align*}
\dMdetjump{2} &= \left[\bar X^{(1)}(\dettime^{-})- \frac{\tilde
F}{\tilde\kappa^{(1)}}\right]-\frac 12\left[\tilde
\kappa^{(1)}\bar X^{(1)}(\dettime^{-})+\tilde \kappa^{(2)}\bar
X^{(2)}(\dettime^{-})-\tilde
F\right]\nonumber\\
  &=\frac{\tilde \kappa^{(2)}}2R(\dettime^{-})- \frac{\tilde
\kappa^{(2)}\tilde
F}{2\tilde
\kappa^{(1)}} ,
\end{align*}
and thus have the distribution
\begin{equation*}
\dMdetjump{2} \sim  \frac{\tilde \kappa^{(2)}}2\left(\RdetJ{2}- \frac{\tilde
F}{\tilde
\kappa^{(1)}}\right) . 
\end{equation*}
where $ \RdetJ{i}  $ is defined as a random variable with distribution equal
to that of the intermotor distance $ \Rdet $ conditioned on the event $ \detmot=i$
that motor $i $ is the one which detaches from the state of both motors attached.
 Using Bayes' rule with Eqs.~\eqref{conddet}-\eqref{probdetone}, we can derive
the probability density function
\begin{equation} \label{rdetj}
p_{\RdetJ{i}} =\frac{p_{R}(r)\dettwoz{i}(r) }{\dettwo{i}}.
\end{equation}
A similar calculation shows that when motor 1 detaches ($\detmot=1 $), then
 $\dMdetjump{1}\sim  \frac{\tilde\kappa^{(1)}}2 (-\RdetJ{1}-  \frac{\tilde
F}{\tilde\kappa^{(2)}})$. In more compact form, the jump in the tracking
variable
when motor $i $ detaches first from the two-motor-attached state has distribution:
\begin{equation}\label{jumpoff}
\dMdetjump{i} \sim   \frac {\tilde \kappa^{(i)}}{2}\left((-1)^{i}\RdetJ{i}-\frac{\tilde
F}{\tilde
\kappa^{(i^{\prime})}}\right),
\end{equation}

\subsubsection{Jumps at attachment of second motor from state of one motor
attached}\label{subsub:attach}
We now compute the statistics of the jumping distances at a time $ \atttime
$ before which only one motor is attached, and at which time the second motor
(with index $ \attmot $) attaches.   Suppose first that we begin with motor
$ 1 $ attached, then motor $2 $ attaches ($\attmot =2 $) at time $ \atttime
$.  

From  Eq.~\eqref{patt}, detached motor
$2$ has a location distributed, conditional on the position of  attached
motor $1$, as
\begin{equation} \label{unbitoz}
\bar X^{(2)}(\atttime^-)\sim  N( \bar X^{(1)}(\atttime^-)-\tilde F_T/\tilde\kappa^{(1)},1/(\tilde\kappa^{(1)}\tilde\kappa^{(2)})).
\end{equation} Under our model that the attaching
motors attaches at a position governed by its detached spatial distribution,
the jump in the central coordinate upon attachment of motor 2 is then
\begin{align}
\dMattjump{2} &=\frac 12\left[\tilde \kappa^{(1)}\bar X^{(1)}(\atttime^-)+\tilde
\kappa^{(2)}\bar X^{(2)}(\atttime^-)-\tilde F_T\right] - 
\left[\bar X^{(1)} (\atttime^-) - \frac{\tilde F_T}{\tilde{\kappa}^{(1)}}\right]
\\
&\sim N\left(0,\frac{\tilde{\kappa}^{(2)}}{4\tilde{\kappa}^{(1)}}\right).
\end{align}
More generally, when motor $i $ is the detached motor which attaches, the
tracking coordinate jumps by an amount
\begin{equation}\label{jumpon}
\dMattjump{i} \sim N\left(0,\frac{\tilde{\kappa}^{(i)}}{4\tilde{\kappa}^{(i^{\prime})}}\right),
\end{equation}
which is always mean zero even when the motors are nonidentical.  We remark
that the reattachment rule of Keller, Berger, et al~\cite{lipowsky2013network},
where the tether
between the
detached motor and cargo is treated as exactly slack (at rest length), and
this motor reattaches similarly at a location where the tether force is zero,
and the cargo instantaneously moves to a position of mechanical force balance,
corresponds to a deterministic version of the rules described for our model
above, using just the means of the random reattachment position and jump
in cargo tracking variable.

\subsection{Effective dynamics in terms of detachment-attachment cycles}\label{sect:procdisp}

Starting
from the fully attached state   $(1,2) $, the motor-cargo
complex will undergo a random number  $\Ncyc $ of \textit{full cycles} between
2-motor
and 1-motor attached states (either $ (1,2) \rightarrow (1) \rightarrow
(1,2) $ or $ (1,2) \rightarrow (2) \rightarrow (1,2) $) and ultimately  a
\textit{terminal cycle}
(either $ (1,2) \rightarrow (1) \rightarrow
 \emptyset $ or $ (1,2) \rightarrow (2) \rightarrow  \emptyset $)  ending
in complete detachment. Thus  $\Ncyc$ is a geometric
random variable with mean $(1-\totdetprob)/\totdetprob$, 
  where the probability of complete
detachment during an initiated cycle
 is defined by
\begin{equation}\label{probcompdet}
\totdetprob= \pdet{2}\frac{\detone{1}}{
\bar a^{ *(2)}+\detone{1}}+(1-\pdet{2})\frac{\detone{2}}{\bar
a^{* (1)}+\detone{2}}.
\end{equation}

For each cycle (either complete or terminal), time advances by a random
increment
\begin{equation} \label{tcycle}
\dTcyc = \dTattb + 
\Delta T_{\detmot{}^{\prime}},
\end{equation}
where $ \detmot{}^{\prime} \equiv 3 - \detmot $ is the index of the motor
remaining attached.  
 
  The distributions
of time $ \dTattb $  spent in the fully attached and time $ \dTdet{i} $ spent
in the state with only motor $i$ attached  are exponentially
distributed random variables with the indicated means:
\begin{equation}\label{timedists}
 \dTattb \sim \Exprand ((\dettwoboth )^{-1}), \quad \dTdet{i} \sim \Exprand
((\bar{a}^{*(i^{\prime})}+\detone{i})^{-1}).
\end{equation}

Similarly, in each cycle the tracking variable will advance by a random increment
\begin{equation}\label{mcycleterms}
\dMcyc = \dMattb +\dMdet{\detmot{}^{\prime}} +\dMattjumpall{\detmot{}}+\dMdetjump{\detmot}.
\end{equation}
Here, we write  $\dMattjumpall{i}$ to extend the random variable $\dMattjump{i}$~\eqref{jumpon}
describing the jump in the cargo tracking variable upon reattachment of motor
$ i$ to omit its contribution (with the value $\dMattjumpall{i} = 0$) in
the terminal cycle when  motor $i^{\prime} $ detaches before motor $ i $
reattaches.
 The distributions for the jump in the cargo tracking variable upon detachment
of motor $i $, $\dMdetjump{i}$,
is given
in Eq.~\eqref{jumpoff}.  By the independence of residence times and the next
state visited in a continuous-time Markov chain, these jumps  are independent
of the time spent in any state of the cycle. On the other hand,
from the results of Subsection~\ref{ssec:homtwomotor} for the effective velocity
and diffusivity in each attachment state,
\begin{align}
 \dMattb|\dTattb &\sim N(\barvtwo \dTattb ,2D_{1,2}\dTattb ), \label{m12cond}\\
\dMdet{i}|\dTdet{i} &\sim N(\Vstate{i} \dTdet{i}, \rho^{(i)}\dTattb ).
 \end{align}  
Like the distributions for $\dTattb$ and $\dTdet{i}$, the displacements 
$ \dMattb $ and $ \dMdet{i} $ within an attachment state are independent
on whether the motor system eventually returns to a two motor attached state.

\section{Effective Transport Characterization}\label{sec:renewal}



For a cooperative system of two motors, we have provided in the
previous section a coarse-grained approximation of the stochastic
process governed by equations (\ref{nondmotors})-(\ref{nondetach}). 
These simplified equations are adequate, under the conditions of
validity of the asymptotic approximations, for computing effective
transport properties of the motor-cargo complex.  We begin in
Subsection~\ref{sect:proc} by computing the processivity measures:  the
mean and variance of the run time and run length.  Then we turn in
Subsection~\ref{sect:velanddiff} to the theoretical calculation of the
effective velocity and diffusivity of the motor-cargo complex.  The proper
definition of these transport statistics is not entirely obvious for a
motor-cargo complex that eventually detaches and terminates progress
along the microtubule.  We discuss two distinct mathematical framings
of velocity and diffusivity in this context, and relate them to
approaches used in analyses of previous models as well as to
experimental approaches.  We then, in turn, compute the velocity and diffusivity
under each of the two mathematical interpretations.

The formulas in these subsections are formulated in terms of  statistics
of the cycles of attachment and detachment presented and derived in Subsection~\ref{sec:cyclestats}.
   In complicated expressions, we will sometimes  have
$ \mu_Y $ denote
the mean, $ \sigma_Y $ denote the standard deviation, and $ \sigma_{Y,Y^{\prime}}
$ denote the  covariance of the random variables $ Y, Y^{\prime} $.

\subsection{Run length and run time statistics} \label{sect:proc}

We now consider the total run time $\Talldet$ and total run length $\Malldet$
taken by an ensemble of a cargo with two cooperative motors before complete
detachment.  For
simplicity, we take the system to start with both motors are attached.
Then $ \Talldet $ is just the first passage time of the coarse-grained
Markov chain from state $ (1,2) $ to the state $ \emptyset $, and $
\Malldet $ is the increment in the cargo tracking variable $ M$ until
absorption at the fully detached state $ \emptyset $.   We may then write
$T$ and $\Malldet$ as random sums of iid random variables $\{\dTcyc^j\}_{j=1}^{\infty}$
and  $\{\dMcyc^j\}_{j=1}^{\infty}$,
where the number of  complete detachment-attachment cycles $\Ncyc$ (and therefore
also the total number of cycles $ \Ncyc + 1 $) has the property of a Markov
(stopping) time for the natural filtration generated by these two sequences
of random variables together with the sequence of Markov chain states visited:
\begin{equation}
T = \sum_{j = 1}^{\Ncyc+1} \dTcyc^j, \quad \Malldet = \sum_{j = 1}^{\Ncyc+1}
 \dMcyc^j.
\end{equation}

This permits us to use Wald's identity (Th. 14.6 of DasGupta~\cite{dasgupta2011probability})
and the second Wald
identity
(Th. 2.4.5 of Ghosh, Mukhopadhyay, and Sen~\cite{ghosh2011sequential})  to
obtain:
\begin{proposition}{(Run length and time  from cycle statistics)} \label{runlengthtimeprop}
\begin{enumerate}
\item
 The mean run time and run length are given by
\begin{align} \label{meanrun}
\mu_{T}  &\equiv \E[T]= \E[\Ncyc+1]\mu_{\dTcyc},\\ \quad 
 \mu_{\Malldet}  &\equiv \E[\Malldet]= \E[\Ncyc+1]\mu_{\dMcyc}. 
\end{align}
\item The variances and covariance of the run time and run length are given
by
\begin{align} \label{varrun}
\sigma^2_T&\equiv \Var(T) = \E[\Ncyc+1]\sigma^2_{\dTcyc}
+ \Var(\Ncyc+1)\mu^2_{\dTcyc},\\
\sigma^2_{\Malldet} &\equiv \Var (\Malldet)=\E[\Ncyc+1]\sigma^2_{\dMcyc}+
\Var(\Ncyc+1) (\mu_{\dMcyc})^2,\nonumber\\
\sigma_{T,\Malldet} &\equiv \Cov(T,\Malldet)=\E[\Ncyc+1]\sigma_{\dTcyc,\dMcyc}+\Var(\Ncyc+1)\mu_{\dTcyc}\mu_{\dMcyc}.
\end{align}

\item The number of complete detachment-attachment cycles $ \Ncyc $ has the
following first and second order statistics:
\begin{align}
\E[\Ncyc+1] &= \frac{1}{\totdetprob}, \\
\Var[\Ncyc+1] &= \frac{1-\totdetprob}{\totdetprob{}^2},
\end{align}
\end{enumerate}
with the probability of complete
detachment during an initiated cycle
given by
\begin{equation}
\totdetprob= \pdet{2}\frac{\detone{1}}{
\bar a^{ *(2)}+\detone{1}}+(1-\pdet{2})\frac{\detone{2}}{\bar
a^{* (1)}+\detone{2}}.
\end{equation} 
\end{proposition}
The statistics of $ \Ncyc $ follow directly from the discussion at the beginning
of Subsection~\ref{sect:procdisp}.
Explicit expressions for the other cycle statistics in Prop. 
\ref{runlengthtimeprop} will be provided
in Subsection~\ref{sec:cyclestats}.  

It may seem surprising that we do not need to subdivide the calculation
into the component from the $ \Ncyc $ complete cycles and the terminal cycle,
since the probability distribution of the attachment jump $ \dMattjumpall{\detmot{}^{\prime}}
$ \emph{does} depend on whether it is a terminal cycle. But Wald's identity
precisely allows us to do this because, viewed jointly, the sequence of 
times taken, cargo motion incurred (including the jump variables), and attachment/detachment
events within each cycle are independent and identically distributed across
cycles.  It is only when one conditions
on the first cycle that leads to complete detachment ($ \Ncyc + 1 $) that
the cargo motion incurred on each cycle is no longer identically distributed.
 Our calculation involving Wald's identity eschews this conditioning step,
which in fact would further complicate the calculation due to its effect
on which motor $ \detmot{}$ detaches during a cycle (see Subsection~\ref{sec:cyclestats}
below).

 \subsection{Effective velocity and diffusion}\label{sect:velanddiff}
The characterization of the effective velocity and diffusivity is not so
straightforward for a motor-cargo complex that eventually detaches from a
microtubule.  One cannot directly take the long-time limit of the ratio of
distance traveled to time, since the motor-cargo complex will detach at a
finite time.  Of course, for cooperative motor models that explicitly model
a rate for reattachment for the motors even from the fully detached state,
one can define an effective velocity and diffusivity in the usual
way, essentially averaging progress over both phases where the motor-cargo
complex is attached or detached from a microtubule~\cite{lipowsky2009review,miles2018analysis,popovic2011compartmental}.
 While such an
effective velocity is meaningful for
characterizing transport, it does not relate so naturally to experimental
observations of particular cargo, which are tracked only while they appear
to be attached to a microtubule.   Moreover, the time until reattachment
could be quite long.  In a model with explicit reattachment from the fully
detached state, one could alternatively and meaningfully define a effective
velocity conditioned on attachment~\cite{klumpp2005cooperative}, but it is
not clear how to similarly define a diffusivity
conditioned on attachment.

In order to describe  the effective velocity and diffusivity
of
cargo \emph{during periods where at least one of its motors is attached 
to a microtubule}, we consider two distinct definitions of effective velocity
and diffusivity
which could be applied to any theoretical or simulation model for motor-cargo
dynamics, without reference to a model for reattachment from a state of complete
detachment:
\begin{enumerate}
\item Pooling  run times $ \Rtime{j} $ and run lengths $ \Rlength{j} $ over
independent experiments $ j=1,\ldots,\Nexp $, and defining the \emph{ensemble}
velocity and diffusivity as:
\begin{align} 
\Vpool &\equiv \lim_{\Nexp\rightarrow \infty} \frac{\sum_{j=1}^{\Nexp} \Rlength{j}}{\sum_{j=1}^{\Nexp}
\Rtime{j}},\label{pooledvel} \\
\Dpool &\equiv \lim_{\Nexp \rightarrow \infty} \frac{\sum_{j=1}^{\Nexp}\left(
\Rlength{j}
- \Vpool \Rtime{j}\right)^2}{2 \sum_{j=1}^{\Nexp} \Rtime{j}}.\label{pooleddiff}
\end{align}
\item 
We may alternatively censor  experiments by requiring that they complete
a certain number  of full cycles, thereby defining the \emph{long-run} velocity
and diffusivity as
\begin{align} 
\Vrun &\equiv \lim_{k\rightarrow \infty}\lim_{\Nexp\rightarrow
\infty} \frac{\sum_{j=1}^{\Nexp} \Rlength{j}\mathbf 1_{\Ncyc^{(j)}>k}}{\sum_{j=1}^{\Nexp}
\Rtime{j}\mathbf 1_{\Ncyc^{(j)}>k}},\label{runvel} \\
\Drun &\equiv \lim_{k\rightarrow \infty}\lim_{\Nexp \rightarrow \infty} \frac{\sum_{j=1}^{\Nexp}\left(
(\Rlength{j}
- \Vrun \Rtime{j})\mathbf 1_{\Ncyc^{(j)}>k}\right)^2}{2 \sum_{j=1}^{\Nexp}
\Rtime{j}\mathbf 1_{\Ncyc^{(j)}>k}}.\label{rundiff}
\end{align}  
\end{enumerate} 
What we have defined as the ensemble velocity should coincide with the velocity
conditioned on attachment in models with explicit reattachment from a completely
detached state~\cite{klumpp2005cooperative}.  With regard to the distinction
in the definition of the long-run transport statistics, 
note that a large number of cycles during a run, $\Ncyc^{(j)}\rightarrow
\infty
$, implies (almost surely) a large run time  $\Rtime{j} \rightarrow \infty
$.  The idea here is that one may often only wish to take data on sufficiently
long runs in experiments, or simulations, in order to downplay transient
effects at the beginning or end of a cargo run, and better characterize the
dynamics in the middle of a run.  Censoring on run time $ \Rtime{j} $ rather
than the number of attachment/detachment cycles $ \Ncyc^{(j)} $ 
in a simulation or experiment would be more natural, but the derivation of
the theoretical expression would be less straightforward, so we leave its
consideration for a later work.  The $ k \rightarrow \infty $ limit is the
more important one in the above definitions as it characterizes a ``long
run''. The $ \Nexp \rightarrow \infty $ limit of many experiments passing
the censoring step is unnecessary for computing the long-run velocity and
only needed for computing the long-run diffusivity since we only take data
on the total time and displacement of a run.   We have left aside here the
capacity, often exploited,
to use observations of the cargo position at intermediate times within a
run to infer velocity or diffusivity through, for example, fitting mean displacement
and mean-squared displacement as a function of time~\cite{surrey2011cin8,king2006skating,andreasson2015mechanochemical}.
More complex statistical definitions of measured velocity
and diffusivity could be accordingly formulated.  We finally remark that
in the limit of small detachment rate, the long-run statistics should converge
to the ensemble transport statistics since most runs will be long~\cite{fricks2013pearson}.

The ensemble definition has good mathematical properties and is an idealized
manner of estimating velocity and diffusivity in experiments,
but we must bear in mind that experiments typically only measure runs of
a cargo that are sufficiently long~\cite{soppina2014superprocessive,visscher1999single,arpag2019kin1kin3,furuta2013measuring},
since short runs are difficult to detect
or disambiguate from noise.  Thus, the actual experimental values might fall
somewhere between the ensemble and long-run definitions given above, so we
will study both.  The selection of runs to record in an experiment
can be censored in other ways as well, for example those with an observable
initial attachment and complete detachment event~\cite{klumpp2008severalk}.
 
Some stochastic simulations  also compute velocity
(and sometimes diffusivity) using the ensemble definitions given above~\cite{epureanu2017superpar},
though some simulation studies compute velocities
and diffusivities in each run, and then average the single-run velocities
and diffusivities over the ensemble~\cite{gross2011arranged,bouzat2016models}
by averaging the ratio of run length to run time ($ \frac{1}{\Nexp} \sum_{j=1}^{\Nexp}
\frac{\Rlength{j}}{\Rtime{j}} $).  The latter
has no inherent connection to long-time properties, but would be approximated
by the long-run definitions if the run times happened to be sufficiently
long in some statistical sense~\cite{fricks2013pearson}.  This velocity estimator,
however, has infinite variance~\cite{fricks2013pearson} while the ensemble
definition~\eqref{pooledvel} (with a large but finite number of samples $
\Nexp $) enjoys the good statistical convergence properties afforded by the
applicability of the central limit theorem.  

We define $ \mathcal A $ as the event that a cycle ends with a return to
the two-motor-attached
state ($(1,2) \rightarrow (i) \rightarrow (1,2)$) rather than to the detachment
of the complex ($(1,2) \rightarrow (i) \rightarrow \emptyset$).  The formulas
for the long-run velocity and diffusivity involve first and second moment
statistics of cycle displacements and durations, conditioned upon $ \mathcal
A $; we denote these conditional statistics by appending ''$ |\mathcal A
$'' in the subscript. 
As $\Ncyc \rightarrow \infty$, the contribution
of the terminal cycles in $\Vrun$ and $\Drun$ become negligible. From the
standard
law of large numbers for independent random variables for the case of ensemble
statistics, and its version for renewal-reward processes~\cite{sr:asp,serfozo2009basics}
for the long-run statistics,
we arrive at the following expressions: 
\begin{proposition}{(Velocity and diffusivity from cycle statistics)} \label{veldiffstats}
If all runs are initialized from a state in which both motors are attached:
\begin{enumerate}
\item The ensemble velocities and diffusivity
are given by
\begin{align}
\Vpool &= \frac{\mu_{\Malldet}}{\mu_{T}} =\frac{\mu_{\dMcyc}}{\mu_{\dTcyc}}
\label{vpool} ,\\
\Dpool &= \frac{\Vpool^2 \sigma^2_{T} + \sigma^2_{\Malldet} - 2 \Vpool \sigma_{T,\Malldet}}{2\mu_T}
\label{dpool} \\
&=\frac{\Vpool^2\sigma^2_{\dTcyc}+ \sigma^2_{\dMcyc} 
-2 \Vpool \sigma_{\dTcyc,\dMcyc}}{2 \mu_{\dTcyc}} \nonumber.
\end{align}
\item The long-run velocity and diffusivity are given by 
\begin{align}
\Vrun &= \frac{\mu_{\dMfull}}{\mu_{\dTfull}}, \label{vrun}\\
\Drun &= \frac 12\left(\frac{\mu_{\dMfull}^{2} \sigma_{\dTfull}^2}{\mu_{\dTfull}^3}+
\frac{ \sigma_{\dMfull}^2}{\mu_{\dTfull}}-\frac{2\mu_{\dMfull}
\sigma_{\dMTfull}}{\mu_{\dTfull}^2}\right)\label{drun}\\
&=\frac{\Vrun^2\sigma^2_{\dTfull}+\sigma^2_{\dMfull}-2\Vrun\sigma_{\dMTfull}}{2\mu_{\dTfull}}.\nonumber
\end{align}
\end{enumerate}
\end{proposition}



We see that the expressions for the ensemble and long-run velocity and diffusivity
in terms of cycle statistics have a similar structure, with the latter involving
a conditioning on the cycle indeed returning to a state of two attached motors
rather than possibly terminating in complete detachment.  
\begin{corollary}
When the two motors in the ensemble have identical parameters, then $ \Vpool
= \Vrun $ and 
\begin{equation*}
 |\Dpool-\Drun| \leq \frac{1}{8\mu_{\dTcyc}} .
 \end{equation*} \label{cor:homo}
 \end{corollary} 
 We prove this corollary in Subsection~\ref{sec:corproof}.  For both kinesin-1/kinesin-1
and kinesin-2/kinesin-2 ensembles which we consider in Section~\ref{sec:sims},
the difference
in diffusivities is less than one percent.  The distinction between ensemble
and long-run transport characteristics therefore appear potentially important
mainly for heterogenous ensembles.

\subsection{Expressions for cycle statistics}\label{sec:cyclestats}
The formulas for the effective motor-cargo transport in Subsections~\ref{sect:proc}
and~\ref{sect:velanddiff} refer to statistics of durations and displacements
within detachment-attachment
cycles.  We now provide formulas for these cycle statistics, followed by
a discussion of how they are derived.

\begin{proposition}{(Explicit expressions of
cycle statistics)} \label{thm:effvelstats}
\begin{enumerate}
\item
The first and second order moments of the unconditional cycle times and displacements
have the explicit forms in terms of the effective nondimensional parameters
defined in the coarse-grained model from Section~\ref{slowswitch}:

\begin{align}
&\mu_{\dTcyc} = \frac 1 {\dettwoboth}   +\frac{\pdet{2}}{\bar{a}^{*(2)}+\detone{1}}+\frac{\pdet{1}}{\bar{a}^{*(1)}+\detone{2}},
\label{propcycstart}\\
&\mu_{\dMcyc} =   \frac{  \barvtwo }{\dettwoboth}+ \frac 12\left(\mu_{\RdetJ{2}}\pdet{2}\tilde\kappa^{(2)}-\mu_{\RdetJ{1}}\pdet{1}\tilde\kappa^{(1)}
  \right)  +\frac{\pdet{2}\Vstate{1}}{\bar
a^{*(2)}+\detone{1}} \label{mumcyc}\\
&+\frac{\pdet{1}\Vstate{2}}{\bar
a^{*(1)}+\detone{2}}-\frac{ \tilde F_T }2\left(\pdet{2}\frac{\tilde
\kappa^{(2)}}{\tilde \kappa^{(1)}}+\pdet{1}\frac{\tilde \kappa^{(1)}}{\tilde
\kappa^{(2)}}\right),\nonumber\\
&\sigma^2_{\dTcyc} =\frac 1 {(\dettwoboth)^2} +\frac{\pdet{2}}{\left(\bar{a}^{*(2)}+\detone{1}\right)^2}+\frac{\pdet{1}}{\left(\bar{a}^{*(1)}+\detone{2}\right)^2}
\label{vartcyc}\\
&+\pdet{2}\pdet{1}\left(\frac{1}{\bar{a}^{*(2)}+\detone{1}}-\frac{1}{\bar{a}^{*(1)}+\detone{2}}\right)^2,
 \nonumber  \\
&\sigma^2_{\dMcyc} =\frac{2D_{1,2}}{\dettwoboth} \mathbb +\left(\frac{\barvtwo}{\dettwoboth}
\right)^2 \label{varmcyc}\\
&+\sum_{i = 1}^2\pdet{i}\left(  \frac{2D_{i'}}{\bar a^{*(i)}+\detone{i'}}
+\left(\frac{\Vstate{i'}}{\bar
a^{*(i)}+\detone{i'}}\right)^2+\frac{1-\totdetprob}{4}\frac{\tilde\kappa^{(i)}
 }{\tilde\kappa^{(i')}} +\frac{1}{4}(\tilde \kappa^{(i)})^2\sigma_{\RdetJ{i}}^2\right)
\nonumber\\
 &+\pdet{1}\pdet{2}\left(\left(\frac{\tilde
\kappa^{(1)}-\tilde
\kappa^{(2)}}{\tilde
\kappa^{(1)}\tilde \kappa^{(2)}}\right)\tilde
F+\sum_{i = 1}^2\left(\frac{(-1)^{i+1}\Vstate{i}}{\bar{a}^{*(i')}+\detone{i}}+\frac{\mu_{\RdetJ{i}}\tilde\kappa^{(i)}}2\right)\right)^2,\nonumber\\
 &\sigma_{\dTcyc, \dMcyc}  = \frac{\barvtwo}{(\dettwoboth)^2}+ \frac{\pdet{2}\Vstate{1}}{(\bar
a^{*(2)}+\detone{1})^2}+ \frac{\pdet{1}\Vstate{2}}{(\bar
a^{*(1)}+\detone{2})^2}  \label{covarmtcyc}\\
&+\pdet{1}\pdet{2}\left(\frac{1}{\bar a^{*(2)}+\detone{1}}-\frac{1}{\bar
a^{*(1)}+\detone{2}}  \right)\nonumber\\
&\times \left( \left(\frac{\tilde
\kappa^{(1)}-\tilde
\kappa^{(2)}}{\tilde
\kappa^{(1)}\tilde \kappa^{(2)}}\right)\tilde
F+\sum_{i = 1}^2\left(\frac{(-1)^{i+1}\Vstate{i}}{\bar{a}^{*(i')}+\detone{i}}+\frac{\mu_{\RdetJ{i}}\tilde\kappa^{(i)}}2\right)\right).
\end{align}

Additional notation used here is $ i'=3-i $ (the index of the ``other'' motor),
the probability $ \pdet{i} = \frac{\dettwo{i}
}{\dettwoboth}$  (\ref{probdetone}) that  motor $i $ detaches first from
the state of both motors attached, the unconditional probability of complete
detachment in a given cycle $ \totdetprob $~\eqref{probcompdet}, and $ \mu_{\RdetJ{i}}
$, the conditional mean of $\Rdet$ given detachment of motor $ i$, which
may be computed from its probability density
(\ref{rdetj}).
\item Expressions of the corresponding cycle statistics which are conditioned
on $\mathcal A$, the cycle being complete and returning to a state of two
motors attached, are the same
as corresponding statistics given in the above equations (\ref{propcycstart})-(\ref{covarmtcyc}),
except that 
\begin{enumerate}
\item In all instances,   $\pdet{i}$ is replaced with
\begin{align}\label{pbayes}
 \pdeta{i}&
=\left(1+ \frac{\bar
a^{(i')}}{\bar a^{(i)}} \frac{\dettwo{i'}}{\dettwo{i}}
\frac{\detone{i'} + \bar a^{(i)}}
{\detone{i}+\bar a^{(i')}}\right)^{-1}.
\end{align}
\item To compute $\sigma_{ \dMcyc|\mathcal A, }$, in equation (\ref{varmcyc})
we replace the term $\totdetprob$ (defined in (\ref{probcompdet})) with 0.
\end{enumerate}
\end{enumerate}
\end{proposition}

\subsubsection{Derivation of unconditional cycle statistics in  Proposition
\ref{thm:effvelstats}}


We begin by computing statistics for run lengths and times for different
attachment
states involved in a cycle.
 For evolution
with two attached motors, we invoke the law of total expectation, conditioning
on  $\dTattb$,  
to obtain
\begin{equation}
\mathbb E[\dMattb ] = \frac{  \barvtwo }{\dettwoboth}.
\end{equation}
  From  (\ref{m12cond}), we may use the law of total
variance, to obtain
\begin{equation}
\Var(\dMattb ) =\frac{2D_{1,2}}{\dettwoboth}  +\left(\frac{\barvtwo}{\dettwoboth}
\right)^2.\end{equation}

Next, from the definition~\eqref{mcycleterms} of $ \dMcyc$ we obtain,
by conditioning on possible values of  $\detmot{}$,   
\begin{align}
& \mathbb \mu_{\dMcyc}= \frac{  \barvtwo }{\dettwoboth}+\pdet{2}\left(\frac{\Vstate{1}}{\bar
a^{*(2)}+\detone{1}}+\frac{\tilde \kappa^{(2)}\mu_{\RdetJ{2}}}2-\frac{\tilde
\kappa^{(2)}\tilde F_T}{2\tilde \kappa^{(1)}} \right)\\&+ \pdet{1}\left(\frac{\Vstate{2}}{\bar
a^{*(1)}+\detone{2}}-\frac{\tilde \kappa^{(1)}\mu_{\RdetJ{1}}}2-\frac{\tilde
\kappa^{(1)}\tilde F_T}{2\tilde \kappa^{(2)}} \right), 
 \end{align}
 which is equivalent to  (\ref{mumcyc}).
A similar argument, referring to Eq.~\eqref{tcycle} and~\eqref{timedists}
yields~\eqref{propcycstart}.

From  (\ref{jumpoff}),   (\ref{jumpon}), and (\ref{mchange}) we can similarly
 compute the following statistics:
\begin{align}
&\E(\dMdet{i}
 )=  \frac{\Vstate{i}}{\bar a^{*(i')}+\detone{i}}, &&   \Var(\dMdet{i} )
= \frac{2D_{i}}{\bar a^{*(i')}+\detone{i}}  +\left(\frac{\Vstate{i}}{\bar
a^{*(i')}+\detone{i}}\right)^2,    \nonumber \\
 &\E(\dMattjump{i} ) = 0, &  &\Var(\dMattjump{i}) 
= \frac{1
 }{4}\frac{\tilde\kappa^{(i)}
 }{\tilde\kappa^{(i')}}, \label{quickformatstart}\\
  &\E(\dMattjumpall{i}) = 0, &  &\Var(\dMattjumpall{i}) 
= \frac{1-\totdetprob
 }{4}\frac{\tilde\kappa^{(i)}
 }{\tilde\kappa^{(i')}}, \label{quickformatend}\\
&\E(\dMdetjump{i})
 = \frac {\tilde \kappa^{(i)}}{2}\left((-1)^{i}\mu_{\RdetJ{i}}-\frac{\tilde
F}{\tilde
\kappa^{(i^{\prime})}}\right), &   &\Var(\dMdetjump{i}) 
=\frac{1
 }{4}(\tilde \kappa^{(i)})^2\sigma_{\RdetJ{i}}^2.  \label{quickformend}
\end{align}
Here, we note that the mean  $\mu_{\RdetJ{i}}$ and variance $\sigma_{\RdetJ{i}}^2$
of the intermotor separation $ \RdetJ{i} $ at the detachment time, given
motor $i $ detaches first,
may be computed
directly from the density given in Eq.~\eqref{rdetj} .
For the special case of constant detachment rates, $\mu_{\RdetJ{i}}=
\mu_R$ and
$ \sigma_{\RdetJ{i}}=\sigma_R $.

To find  $\sigma^2_{\dTcyc}  $ and $\sigma^2_{\dMcyc} $, observe that
the times spent  and dynamics within the fully attached and one-motor detached
states are independent,  which implies\begin{align}\label{tcvar}
\sigma^2_{\dTcyc}  &= \Var (\dTattb )
+ \Var( \dTdet{\detmot{}^{\prime}}) , \\
\sigma^2_{\dMcyc} &= \Var(\dMattb )
\nonumber
+\Var(\dMdet{\detmot{}^{\prime}} +\dMattjumpall{\detmot}+\dMdetjump{\detmot}).
\nonumber
\end{align}
From (\ref{timedists}), $\Var(\dTattb )= 1/(\dettwoboth
)^2$.
For the second term in (\ref{tcvar}), we  use the law of total variance,
conditioning
on $ \detmot$, to obtain 
\begin{align}
\Var( \dTdet{\detmot{}^{\prime}})
= \Var ( \dTdet{1} )\pdet{2}+\Var ( \dTdet{2} )\pdet{1}+\left(\E[ \dTdet{1}
- \dTdet{2} ]\right)^2\pdet{2}\pdet{1},\nonumber
\end{align}
which, using (\ref{timedists}), yields (\ref{vartcyc}). A similar calculation
gives (\ref{varmcyc}), with attachment and detachment jumps, due to their
association with the fast dynamics of the cargo and detached motors, taken
as independent of each other and of the progress of the cargo tracking variable
during the time one motor was attached.


It remains to calculate the covariance $\sigma_{\dTcyc, \dMcyc}$, which
may be decomposed as the sum  
 \begin{align} 
&\sigma_{\dTcyc, \dMcyc}= \mathrm{Cov}(\dMattb, \dTattb)  \\
&+ \mathrm{Cov}(\dMdet{\detmot{}^{\prime}} +\dMattjumpall{\detmot}+\dMdetjump{\detmot},
\dTdet{\detmot{}^{\prime}})
\end{align}
The first term follows easily from the law of total covariance, conditioning
on $ \dTattb $:
\begin{align}
&\mathrm{Cov}(\dTattb,\dMattb ) = \frac{\barvtwo}{(\dettwoboth)^2}, 
\end{align}
Through another application of the law of total covariance, conditioning
on $\detmot$ and noting the conditional independence of the jumps in
the cargo tracking variable $ \dMattjumpall{\detmot} $ and $\dMdetjump{\detmot}
$ from the residence times $ \dTdet{\detmot{}^{\prime}} $,  we may write
\begin{align*}
& \mathrm{Cov}(\dMdet{\detmot{}^{\prime}} +\dMattjumpall{\detmot}+\dMdetjump{\detmot},
\dTdet{\detmot{}^{\prime}}) \\
&= \pdet{2}\mathrm{Cov}(\dMdet{1} ,\dTdet{1})+\pdet{1}\mathrm{Cov}(\dMdet{2}
,\dTdet{2})\\
&+\pdet{1}\pdet{2}\left(\E[\dTdet{1}]-\E[\dTdet{2}]  \right)\\
&\times (\E[\dMdet{1}+\dMattjumpall{2}+\dMdetjump{2}]-\E[\dMdet{2}+\dMattjumpall{1}+\dMdetjump{1}]).
\end{align*}
A direct calculation of each of these terms    yields (\ref{covarmtcyc}).

\subsubsection{Derivation of conditional cycle statistics in  Proposition
\ref{thm:effvelstats}}

 By the independence of residence time of a state and
the next state in a continuous-time Markov chain, the conditioning upon re-entry
into the state of two attached
motors does not affect distributions
of attachment and detachment times given in (\ref{timedists}), nor the components
of the cargo-tracking displacements
$\dMattb$, $ \dMdet{i} $, $\dMattjump{i}$, and $ \dMdetjump{i}$.  What is
affected is the probability distribution of which motor is the one to detach
from the state of two attached motors.
 
Let $\detmotfull{}$ 
denote the  index of the motor which detaches during a cycle, conditioned
on the event $ \mathcal A $ of next returning to a two-motor attached state
rather than to a state of complete detachment.  
We can compute the distribution of $\detmotfull{}$
 through Bayes' rule, with
 \begin{align*}
 \pdeta{i}&
 :=  \mathbb P(\detmot=i|\mathcal A )
\\&= \mathbb P(\mathcal A |\detmot=i)\frac{
\mathbb P(\detmot=i)}{ \mathbb P(\mathcal A)}
 = \frac{\frac{\bar a^{(i)}}{\detone{i'}+\bar a^{(i)}}\left(\frac{\dettwo{i}}{\dettwoboth}\right)}{\frac{\bar
a^{(i)}}{\detone{i'}+\bar a^{(i)}}\left(\frac{\dettwo{i}}{\dettwoboth}\right)+
\frac{\bar
a^{(i')}}{\detone{i}+\bar a^{(i')}}\left(\frac{\dettwo{i'}}{\dettwoboth}\right)}
\\
 &=\left(1+ \frac{\bar
a^{(i')}}{\bar a^{(i)}} \frac{\dettwo{i'}}{\dettwo{i}}
\frac{\detone{i'} + \bar a^{(i)}}
{\detone{i}+\bar a^{(i')}}\right)^{-1}.
\end{align*}
Note the conditioning on returning to the state of two attached motors biases
the probability distribution for which motor detaches toward the one that
is more likely to reattach.

When conditionining $\dMcyc$
on the event $\mathcal A$, the only variables affected are $\detmot$ and
$\dMattjumpall{\detmot}$.  Thus, expressions in (\ref{propcycstart})-(\ref{covarmtcyc})
now are computed with $\mathbb{P} (\detmot = i| \mathcal A) =\pdeta{i}$ rather
than $\pdet{i}$.  To obtain $\sigma_{ \dMcyc|\mathcal A, }$ we
note that $ \dMattjumpall{\detmot}|\mathcal
A \sim \dMattjump{\detmotfull} $, and carry out calculations similar to those
which yield (\ref{varmcyc}).
 
\subsubsection{Derivation of special case of identical motors in Corollary~\ref{cor:homo}}
\label{sec:corproof}

  From these formulas, we observe that if two motors in an ensemble have
identical parameters, it follows from symmetry that $\pdet{1} =\pdeta{1}
= 1/2$, and consequently   $\Vpool = \Vrun$.
 Effective diffusivities $\Dpool$ and $\Drun$ for identical motors ensembles
 are,
in general, not equal due to the difference in the term involving $ \totdetprob$
 in equation $(\ref{varmcyc})$.
However, a straightforward estimate comparing the effect of this term on
 (\ref{dpool}) and (\ref{drun})
shows  that the diffusivities differ at most by $1/(8\mu_{\dTcyc})$.

\section{Simulations}\label{sec:sims}

\begin{figure}
 \includegraphics[width=\linewidth]{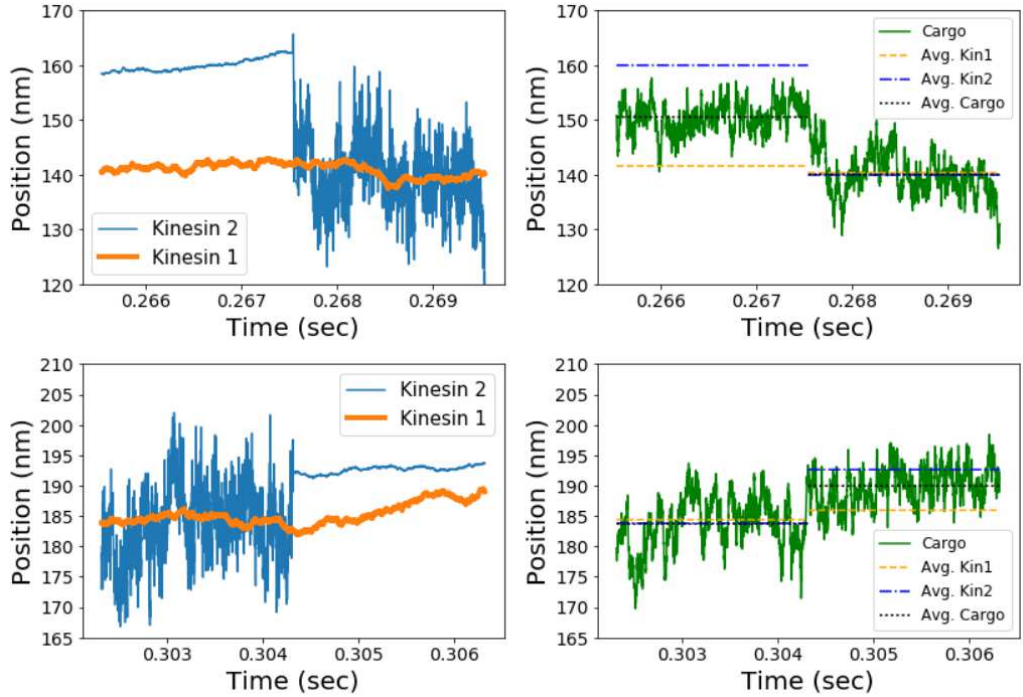}
\caption{Switching behavior for a sample path of the kinesin-1/kinesin-2
complex simulation under trap force $ F_T =0 $ using double exponential detachment
rate functions and
parameters
taken from Table~\ref{paramtable}. Recall that the
molecular motor dynamics are approximated via continuous
diffusion processes, so the small fluctuations ($ <8$ nm) of attached motors
are a model artifact that are not supposed to affect the central concerns
of this work. Top left: Motor behavior at a detachment
event, with kinesin-1 (thick orange) attached at all times shown, and kinesin-2
(thin blue) detaching near  $t = .2675$ s.  Top right: Behavior of cargo
during this detachment event. Dashed lines denote numerical averages  for
positions of motors and cargo taken over $.002$ seconds before and after
detachment. 
  Bottom left: Motor behavior at an attachment
event, with kinesin-1 (thick orange)
attached at all times shown, and kinesin-2 (thin blue) attaching near  $t
= .3043$ s. Bottom right: Behavior of
cargo
during this attachment event. Dashed lines denote numerical averages  for
positions of motors and cargo taken over $.002$ seconds before and after
attachment. } 
\label{fig:simjump}
\end{figure}

In this section, we compare  theoretical and sample statistics  through direct
simulation of equations (\ref{motors})-(\ref{driftu})  for two motor ensembles.
Both homogeneous
(kinesin-1/kinesin-1 or kinesin-2/kinesin-2) and heterogeneous (kinesin-1/kinesin-2)
ensembles are simulated,  using  the parameters
in Table \ref{paramtable}.
 For each ensemble, we  consider optical trap forces of $F_T = -5, 0, $and
$5$ pN.  We also considered two separate detachment models. The first  utilizes
the double exponential function given in (\ref{demodel}).
For comparison, we also use a constant detachment rate set equal
to the average of the double-exponential detachment rate model against the
stationary distribution of the force $ F $ applied by the cargo when the
motor in question is the only one attached with no trap force.  Therefore,
at zero trap force, both models have
the same nondimensional effective detachment rates $ \detone{i}
$, but the first model has a double exponential function  describing the
detachment rate as a function of force while the second model has a force-independent
detachment rate.  Note that, for the constant detachment rate case, this
would also be the detachment rate  from the state with two motors attached,
but would in general differ from the effective detachment rate $ \dettwo{i}
$~\eqref{deffmotori} for the double-exponential detachment rate model.  The
effective detachment rates from the state of one motor attached will be higher
under assisting or opposing trap forces for the double-exponential detachment
rate model relative to the constant detachment rate model.
In all cases, we used constant values $a^{(i)}$ for attachment rates.
 Values of parameters related to attachment and detachment are found in
Table \ref{paramtable}.  

The stochastic differential equations~(\ref{motors})-(\ref{cargo})
were simulated by an Euler-Maruyama discretization with a time increment
$\Delta
t = 10^{-6}$s. The random switching was discretized with respect to the
same time interval, with probabilities to switch as the momentary rate multiplied
by the time step. 
For stability and accuracy concerns,  the time step was
selected to be less than the time scale $\gamma_m/\kappa$ for the drift of
unattached motors (and consequently the larger time scales for attached motors,
cargo, and switching dynamics).    Our choice of the nondimensional force-velocity
curve $g$ in Eq.~\eqref{driftu}
is the same
used in McKinley, Athreya, et al~\cite{mckinley2012asymptotic},
defined by 
\begin{equation}
g(x) = A-B\tanh(Cx-D), \label{eq:fv}
\end{equation}
with $g(x) \rightarrow 1.2$ as $x \rightarrow -\infty$ and $g(x) \rightarrow-.1$
as $x\rightarrow  \infty$.  With the requirements $g(0) = 1$ and $g(1)
= 0$, the constants $A,B,C$, and $D$ may be uniquely determined numerically.

We show a sample path of motor and cargo positions in Fig.~\ref{fig:simjump}
near times of attachment and detachment at zero trap force $ F_T =0 $.  As
predicted by the calculation~\eqref{mdef}
(or the force balance analysis of~\cite{klumpp2005cooperative,klumpp2015review}),
the mean cargo position is a weighted average of the position of kinesin-1
and kinesin-2 when both motors are attached. When one motor is attached,
 the detached motor and cargo both have mean position equal to the current
position of the attached motor.
Fig.~\ref{fig:simjump} also suggests that the transient period for motors
and cargo to change their relative positions is small compared to  times
between attachment or detachment.     In particular, we can interpret the
jumps in the cargo position at motor detachment events as potentially corresponding
to ``flyback'' seen in experimental traces~\cite{trybus2013pauses,mallik2018coin,feng2018motor}.

\begin{table}
\resizebox{\columnwidth}{!}{
\begin{tabular}{|c||c|c||c|c||c|c|} \hline 

 &  \multicolumn{2}{|c||}{$F_T = -5$ pN }   &\multicolumn{2}{|c||}{$F_T =
0$ pN
}    &\multicolumn{2}{|c|}{$F_T = 5$ pN}\\\hline
 & \textbf{Simulation}&  \textbf{Theory} &\textbf{Simulation}  & \textbf{Theory}&
 \textbf{Simulation} &\textbf{Theory }\\ \hline
  & \multicolumn{6}{|c|}{\textbf{kinesin-1/kinesin-1}} \\ \hline
 $\Vrun$ &$951\pm 4$& 946 &$758\pm4 $&755
&$257\pm5$ & 257 \\
\hline
$\Vpool$ & $951\pm 3$& 946&$758\pm 2$ &755 &$256\pm 3$&
257\\\hline
$\Drun$ & $2700\pm200$ & 2500&$2100\pm200$&$2100$ &$3900
\pm 300$&
3400\\\hline
$\Dpool$ &$2500 \pm 100$& 2500 &$2100\pm100 $&2100
&$3700\pm200$ & 3400\\
\hline
$\E[\Ncyc+1]$ & $1.72\pm .03$& 1.74&$1.73\pm .03$ &1.74 &$1.74
\pm .03$&
1.74\\\hline
$\E[ \Rtimegen]      $ & $.34\pm.01$ & .33&$.35\pm.01$&$.35$ &$.34 \pm .01$&
.35\\\hline
$\E[\Rlengthgen]$ & $326\pm6$ & 329&$263\pm5$&$263$ &$88 \pm 2$&
90\\\hline  
 & \multicolumn{6}{|c|}{\textbf{kinesin 2/kinesin 2}} \\\hline
 $\Vrun$ &$628\pm 2$& 632 &$483\pm2 $&482
&$168\pm3$ & 174 \\
\hline
$\Vpool$ & $630\pm 1$& 632&$483\pm 1$ &482 &$171
\pm 1$&
174\\\hline
$\Drun$ & $900\pm100$ & 1000&$610\pm60$&$630$ &$1600\pm
100$&
1200\\\hline
$\Dpool$ &$950\pm 40$& 1000 &$630\pm30 $&620
&$1350 \pm60$ & 1220\\
\hline
$\E[ \Ncyc+1]$ & $4.36\pm .08$& 4.29&$4.24\pm .08$ &4.29 &$4.40\pm .09$&
4.29\\\hline
$\E[ \Rtimegen] $ & $.66\pm.01$ & .65&$.64\pm.01$&$.65$ &$.67\pm .01$&
.65\\\hline
$\E[ \Rlengthgen]$ & $411\pm9$ & 408&$311\pm6$&$311$ &$114 \pm 3$&
112\\\hline
 & \multicolumn{6}{|c|}{\textbf{kinesin-1/kinesin-2}} \\\hline
 $\Vrun$ &$749\pm 6$& 727 &$583\pm4 $&578
&$202 \pm5$ & 210 \\
\hline
$\Vpool$ & $728\pm 3$& 688&$569\pm 2$ &554 &$200\pm
2$&
198\\\hline
$\Drun$ & $4000\pm400$ & 4400&$1800\pm100$&$2200$ &$2800
\pm 300$&
2200\\\hline
$\Dpool$ &$3500 \pm 200$& 3700 &$1800\pm100 $&2000
&$2300 \pm100$ & 2200\\
\hline
$\E[ \Ncyc+1]$ & $2.36\pm .04$& 2.45&$2.44\pm .04$ &2.45 &$2.44
\pm .04$&
2.45\\\hline
$\E[ \Rtimegen] $ & $.42\pm.01$ & .42&$.44\pm.01$&$.42$ &$.44 \pm .01$&
.42\\\hline
$\E [\Rlengthgen]$ & $308\pm6$ & 305&$250\pm5$&$242$ &$87 \pm 2$&
88 \\\hline
\end{tabular}
}
\caption{\textbf{Simulations  with
constant detachment rate model}.  The detachment rates $ d^{(i)} (F) $ are
taken
to be constants obtained by averaging the double exponential detachment rate
model~\eqref{demodel} against the stationary distribution of the force $
F $ when only the motor
in question is attached.  The columns are organized by applied trap force
$ F_T $, with positive (negative) values corresponding to hindering (assisting)
forces.  The theoretical values are computed according to the formulas from
Section~\ref{sec:renewal} while the simulated values are obtained from 2,000
Monte Carlo simulations conducted as described in Section~\ref{sec:sims}.
Units of time and distance are measured
in seconds and nanometers, respectively.
The means of  for the number of cycles (including the terminal one) $\Ncyc
+1$, run time $\Rtimegen$,
and run length $\Rlengthgen$ are estimated with the sample mean,
where intervals denote
the standard error.  The errors in the ensemble and long-run
velocities ($\Vpool$ and $\Vrun$) and diffusivities ($\Dpool$ and $\Drun$)
are obtained
through bootstrap sampling with 1,000 bootstrap samples.} \label{resultsconst}
\end{table}

\begin{table}
\resizebox{\columnwidth}{!}{
\begin{tabular}{|c||c|c||c|c||c|c|} \hline 

 &  \multicolumn{2}{|c||}{$F_T = -5$ pN }   &\multicolumn{2}{|c||}{$F_T =
0$ pN
}    &\multicolumn{2}{|c|}{$F_T = 5$ pN}\\\hline
 & \textbf{Simulation}&  \textbf{Theory} &\textbf{Simulation}  & \textbf{Theory}&
 \textbf{Simulation} &\textbf{Theory }\\ \hline
  & \multicolumn{6}{|c|}{\textbf{kinesin-1/kinesin-1}} \\ \hline
 $\Vrun$ &$1021\pm7 $& 1026 &$780\pm4 $&776
&$291 \pm5$ & 289 \\
\hline
$\Vpool$ & $1018\pm 4$& 1026&$785\pm 3$ &776 &$296\pm 4$&
289\\\hline
$\Drun$ & $2800\pm200$ & 3000&$2300\pm200$&$2200$ &$3900
\pm 200$&
3500\\\hline
$\Dpool$ &$2700 \pm 100$& 3000 &$2200\pm100 $&2200
&$3700 \pm100$ & 3500 \\
\hline
$\E[\Ncyc+1]$ & $1.31 \pm.01$& 1.30&$1.74\pm .03$ &1.74 &$1.45
\pm .02$&
1.44\\\hline
$\E [\Rtimegen] $ & $.15\pm.01$ & .14&$.33\pm.01$&$.32$ &$.30 \pm .01$&
.28\\\hline
$\E [\Rlengthgen]$ & $148\pm3$ & 144&$262\pm5$&$247$ &$88\pm 2$&
80\\\hline  
 & \multicolumn{6}{|c|}{\textbf{kinesin-2/kinesin-2}} \\\hline
 $\Vrun$ &$675\pm3$& 675 &$487 \pm2 $&482
&$19\pm8$ & 67 \\
\hline
$\Vpool$ & $676 \pm2$& 675&$487\pm1$ &482 &$37\pm5$&
67\\\hline
$\Drun$ & $1300 \pm100$ & 1300&$630\pm60$&$690$ &$2600 \pm200$&
1700\\\hline
$\Dpool$ &$1280 \pm50$& 1260 &$650\pm30 $&680
&$2200 \pm100$ & 1600\\
\hline
$\E [\Ncyc+1]$ & $2.67\pm.05$& 2.75&$4.40 \pm.09$ &4.28 &$1.45\pm.02$&
1.47\\\hline
$\E [\Rtimegen] $ & $ .29\pm.01$ & .30&$.63\pm.01$&$.57$ &$.10 \pm.01$&
.10\\\hline
$\E [\Rlengthgen]$ & $197\pm4$ & 201&$308\pm7$&$276$ &$3.6\pm.5$&
6.7\\\hline
 & \multicolumn{6}{|c|}{\textbf{kinesin-1/kinesin-2}} \\\hline
 $\Vrun$ &$812\pm 6$& 878 &$615\pm4 $&612
&$196\pm6$ & 208 \\
\hline
$\Vpool$ & $804\pm4 $& 823&$604\pm 2$ &586 &$187
\pm 4$&
185\\
\hline
$\Drun$ & $2800\pm200$ & 5800&$2200\pm200$&$2600$ &$2600\pm
200$&
2200\\\hline
$\Dpool$ &$3100 \pm 100$& 5500 &$2300\pm100 $&2500
&$2600 \pm100$ & 2300\\
\hline
$\E [\Ncyc+1]$ & $1.77\pm .03$& 1.89&$2.81\pm .05$ &2.73 &$1.77
\pm .02$&
1.76\\\hline
$\E[\Rtimegen] $ & $.21\pm.01$ & .20&$.49\pm.01$&$.42$ &$.20 \pm .01$&
.20\\\hline
$\E [\Rlengthgen]$ & $169\pm3$ & 178&$293\pm6$&$258$ &$38 \pm 1$&
42\\\hline  
\end{tabular}
}
\caption{\textbf{Simulations  with
double exponential detachment rate model}. The detachment rates $ d^{(i)}
(F)
$ are given by the double exponential detachment rate model~\eqref{demodel}.
The columns are organized by applied trap force $ F_T $, with positive (negative)
values corresponding to hindering (assisting) forces.  The theoretical values
are computed according to the formulas from Section~\ref{sec:renewal} while
the simulated values are obtained from 2000 Monte Carlo simulations conducted
as described in Section~\ref{sec:sims}.
Units of time and distance are measured
in seconds and nanometers, respectively.
The means of  for the number of cycles (including the terminal one) $\Ncyc
+1$, run time $\Rtimegen$,
and run length $\Rlengthgen$ are estimated with the sample mean,
where intervals denote
the standard error.  The errors in the ensemble and long-run
velocities ($\Vpool$ and $\Vrun$) and diffusivities ($\Dpool$ and $\Drun$)
are obtained
through bootstrap sampling with 1,000 bootstrap samples.
} \label{resultsde}
\end{table}

For each combination of motor ensemble (kinesin-1/kinesin-1, kinesin-2/kinesin-2,
or kinesin-1/kinesin-2), detachment model (constant or double exponential),
and trap force strength
($F_T = -5,0,$ or 5 pN), we simulated $S = 2,000$  runs, each  beginning
with two attached
motors and cargo at identical positions along the microtubule, and terminating
with complete detachment from the microtubule. All experiments have the same
initial conditions  $X_1(0) = X_2(0) = Z(0) = 0$.  However, due to  repositioning
from force balance, we should expect the cargo to quickly readjust to $10$
nm for $F_T = -5$ pN and to $-10$ nm for $F_T = 5$ pN.  We include these
corrections in reporting our simulation results.
Table \ref{resultsconst} reports these simulation results for constant
detachment rates and Table \ref{resultsde} for double exponential detachment
rate functions.  For the means of run times $\Rtimegen$, run lengths $\Rlengthgen$,
and number
of cycles $\Ncyc$, estimates are given by sample means with intervals of
the  standard error. The ensemble velocities and diffusions are estimated
using the finite $ S=2,000$  version of the formulas~\eqref{pooledvel} and~\eqref{pooleddiff},
with errors estimated from bootstrapping.  
Specifically,  from
our data   $\Theta
= (\Rtime{j},\Rlength{j})_{j = 1, \dots, S}$ for the run times and run lengths
in the $ S=2,000$ simulations,  we drew $B = 1,000$ 
bootstrap samples $\Theta^*_b = (\Rtime{j}_b , \Rlength{j}_b)_{j = 1, \dots,
S}$, for each $b = 1
, \dots, B$, where each  $(\Rtime{j}_b , \Rlength{j}_b) $ denotes a random
sample
\textit{with replacement}  from  $\Theta$.  
For  $\Vrun$~\eqref{runvel} and  $\Drun$~\eqref{rundiff}, we use a similar
procedure, with now the 
dataset $\Theta$ thinned to those $ S/10 = 200 $ data pairs associated to
the run lengths $ \Rlength{j} $ in the top decile. 

Our main concern here will be the question of adequacy of the theoretical
results based on the asymptotic analysis relative to the Monte Carlo simulations,
but we first remark on how the velocities, both theoretical and simulated,
can be quite a bit faster than either of the kinesin-1 or kinesin-2 maximum
speeds  when the trap force is assisting ($F_T = -5 \mathrm{pN}$).  Note
first that
under our force-velocity model~\eqref{eq:fv}, the motors can move 20\% faster
than
their unloaded speeds, listed in Table~\ref{paramtable}, under assisting
loads.  The
reason the reported effective velocities can exceed even this figure is the
contribution from the jumps in the cargo tracking variable at detachment
events (Subsubsection~\ref{subsub:detach}).  Indeed, the cargo with no motors
bound will travel under an assisting
force at an average speed $ F_T/\gamma \sim 5 \times 10^5 \mathrm{nm}/\mathrm{s}
$.  Of course we are only considering the cargo during times where one of
the associated motors is attached to a microtubule, but this indicates the
cargo can move considerably more quickly than the motor speeds when one motor
detaches.  So what is really causing these large velocities under assisting
forces is typically that the lagging motor (under stronger force and therefore
higher detachment rate since the cargo is typically \emph{ahead} of the motors
under assisting force) detaches, freeing the cargo to quickly move forward
to the new quasi-equilibrium with the remaining attached motor, and meanwhile
the detached motor also quickly moves up to the cargo's position and reattaches
near the cargo's new position, becoming now the leading motor.  These jumpy
adjustments can allow the cargo to move at a large velocity, at least until
the cycle is broken by the attached motor detaching before the detached motor
reattaches.

Returning to the central question of the adequacy of the theoretical asymptotic
approximation, we see from Table~\ref{resultsconst} that the simulations
well support the theoretical approximations for models with constant detachment
rates.  The diffusivities are somewhat underestimated for hindering trap
forces, and overestimated for the kinesin-1/kinesin-2 ensemble at zero trap
force.  These issues carry over to the double exponential detachment rate
model in Table~\ref{resultsde}, with now a substantial overestimation of
diffusivity for the kinesin-1/kinesin-2 ensemble with assisting trap force.
  These discrepancies can be traced to order of magnitude
errors in some of the second moment cycle statistics (Eqs.~\eqref{vartcyc}
through~\eqref{covarmtcyc}), which are apparently more sensitive to the non-ideal
scale separation.

A more fundamental discrepancy emerges for the kinesin-2/kinesin-2 ensemble
with hindering trap force, where the mean run length but not the mean run
time is  overestimated by a factor of two by the theory, and the velocity
similarly overestimated.
What appears to make this case the most problematic for the theory is the
failure of the assumption that detachment from  the state (1,2) with both
motors attached takes place on a time scale long compared with that required
for the intermotor separation $ R$ to reach its stationary distribution~\eqref{statr}.
First of all, the nondimensional effective rate of detachment from this state,
$ \dettwoboth =  0.24 $, is the highest for this case out of all considered,
and is therefore the least well separated from the $ \ord(1) $ time scale
of the relaxation dynamics of the intermotor separation $ R$.  Moreover,
the force scale of detachment under hindering forces (for both kinesin-1
and kinesin-2) is about $ F_{d+} = 2 \mathrm{ pN}$ (Table~\ref{paramtable}),
which is on the order of the  load carried by each motor under a trap force
of 5 pN.  Upon reattachment from the state of one attached motor, the leading
motor will still be carrying approximately 5 pN of load, since the recently
reattached motor is typically near the cargo position and relaxed, and so
the leading motor will be considerably more likely to detach shortly after
reattachment than it would under an averaged theory where both motors are
carrying 2.5 pN of load on average.   Kinesin-1 is similarly sensitive to
fluctuations in hindering load, but its lower base rate of detachment $ d_{0+}$
appears not to lead to a substantial violation of the scale separation assumption
(Table~\ref{paramtable}).

\section{Discussion and Conclusions} \label{sec:conc}

\subsection{Summary and Related Work}
We have developed and analyzed a mathematical model for the transport of
cargo by multiple, nonidentical molecular motors  along a microtubule.  The
spatial dynamics are formulated in terms of stochastic differential equations,
coarse-graining implicitly over the stepping dynamics of the motors.  The
process of detachment is modeled via a Cox process, in that the detachment
rate depends on the spatial configuration of the motor-cargo complex, which
in turn is a random process governed by the stochastic differential equations.
 Nondimensionalization revealed an at least nominal separation of time scales
between detached motor dynamics, cargo dynamics, attached motor dynamics,
and attachment/detachment processes.  For the case of two motors attached
to a cargo, we exploited this scale separation by successive averaging and
homogenization procedures to arrive at an effective continuous-time Markov
chain for the attachment states of the two motors, together with random displacements
of a cargo tracking variable associated with each visit to a state.  
The cargo tracking variable is just a smoothed representation of the position
of the cargo that has the same long-time dynamics.  The displacements of
this cargo tracking variable in each state also include jumps associated
with attachment and detachment events where the cargo tracking variable adjusts,
on a fast time scale, to the new state.
We developed analytical formulas for the effective velocity, diffusivity,
and processivity of the cargo by an application of the law of large numbers
and renewal-reward asymptotics to a decomposition of the coarse-grained Markov
chain into regeneration cycles.  \footnote{This latter procedure is presented
in more
generality in a separate publication ``Renewal reward perspective on linear
switching diffusion systems in models of intracellular transport''  by M.~V~Ciocanel,
J.~Fricks, P.~R.~Kramer, and S.~A.~McKinley}.  

Miles, Lawley, and Keener~\cite{miles2018analysis} previously pursued in
a similar spirit an analysis
of effective transport and processivity of a cargo with multiple motors attached
via renewal reward theory.  Their procedure, as ours, relies on a separation
of time scales between the continuous dynamics of spatial motion of the motors
and cargo and the attachment and detachment kinetics.  Our use of the cargo
tracking variable in Subsection~\ref{ssec:track} to provide a representation
of the cargo position even when it has been explicitly removed as a fast
variable is similar in spirit to the study in~\cite{miles2018analysis} of
the conditional expectation of the cargo (and motor) variables given the
attachment/detachment state of the motors.  While our methods of analysis
share these similarities with~\cite{miles2018analysis}, our model and analysis
does offer several complements.  First of all, we remark that~\cite{miles2018analysis}
start with a stepping model for the motors, while we opted for a coarse-grained
stochastic differential equation to unify the mathematical description with
the cargo dynamics.  We note below that our approach could be extended to
motor stepping models, and we intend to do this in future work.  Our model,
on the other hand, has the following distinctive features relative to the
model of~\cite{miles2018analysis}:  First, we allow the motors to be of different
types, since we are interested in understanding dynamics of heterogenous
ensembles as studied for example in Feng, Mickolajczyk, et al~\cite{feng2018motor}.
 \cite{miles2018analysis}
was rather motivated to explain the emergence of transport ability of homogenous
collections of the non-processive motor Ncd~\cite{furuta2013measuring}. 
Secondly,  we model motor dynamics and detachment rates in terms of the instantaneous
force felt and thus the spatial configuration.  \cite{miles2018analysis}
 rather model motor dynamics and detachment  in terms of the number of currently
attached motors, which could be viewed as a phenomenological way of accounting
for the different distributions of forces experienced, but this framework
does not adapt well to mixed motor types.
Third, we provide a statistical model for the detached motors and where they
reattach, rather than assume they are always exactly at the current cargo
location.  Finally, though we focused on the same kind of linear spring model
for the motor-cargo tether as was adopted in the model of~\cite{miles2018analysis},
our mathematical framework can handle nonlinear tether models (Appendix~\ref{app:twomotorapp}).
 This is important for comparison with experiment, as we spell out below.

With regard to the mathematical analysis of the models, our coarse-grained
velocities, diffusivities, and detachment rates within each attachment state
reported in Section~\ref{slowswitch} involve more complex formulas as we
have attempted to explicitly model force dependence of the dynamics.  This
accounting for the effects of the spatial distribution of the motors also
is shown in Subsection~\ref{subsub:effdet} to lead to jump contributions
of the effective cargo position, with generically nonzero means, when motors
attach and detach and the cargo adjusts to the new statistical quasi-equilibrium.
  Effective velocity is computed in~\cite{miles2018analysis} through renewal-reward
asymptotics applied to a long time over which the cargo is allowed to go
through periods of having no motors attached to the microtubule; the results
can be renormalized in the usual way to estimate the average velocity while
attached.  We have eschewed this setup of allowing reattachment from a completely
detached state because it does not allow for a computation of effective diffusivity
during a run in which the cargo is attached to a microtubule in the same
way it does for effective velocity.  Rather we contemplated two asymptotic
idealizations of experiments or simulations which terminate when the cargo
disassociates from a microtubule:  a large ensemble of runs, or a collection
of runs which are censored to be sufficiently long.  Either setup permits
the computation of effective velocity and effective diffusivity through a
law of large numbers, based on the random increments of time and cargo displacement
during a regeneration cycle of detachment and reattachment to the fully attached
state, terminated by a cycle of complete detachment.  The effective velocity
and diffusivity differ somewhat when computed under the two setups, primarily
because the terminal cycle of complete detachment is treated as negligible
when only long runs are considered.  For homogenous ensembles, as shown in
Corollary~\ref{cor:homo}, the effective velocities are in fact identical
and the effective diffusivities rigorously close.  For kinesin-1/kinesin-2
ensembles, the predictions of effective velocity and effective diffusivity
are found in Tables~\ref{resultsconst} and~\ref{resultsde} to differ under
the two scenarios by only a few percent in almost all cases.

\subsection{Conceptual Findings from Mathematical Results}
Inspection of the effective transport formulas presented in Sections~\ref{slowswitch}
and~\ref{sec:renewal} shows (once redimensionalized) that they do not depend
on the friction coefficients $ \gamma $ and $ \gamma_m^{(i)} $ for the cargo
and detached motor dynamics, respectively.  These coefficients only need
to be small enough that the cargo and detached motor dynamics are indeed
fast relative to the attached motor dynamics so that our separation of scales
arguments are valid; then their precise values are not relevant.  On the
other hand, the properties of the motor velocities and diffusivities while
attached 
play
clear roles in the determination of associated statistics for two motor systems.
 The tether spring constant $ \kappa^{(i)} $ plays an important role in setting
the force scale of thermal fluctuations $ \sqrt{k_B T \kappa^{(i)}} $, whose
ratio to stall force (in the nondimensional parameter $ s^{(i)} $) and to
detachment force scale (in the nondimensional parameter $ u^{(i)} $) potentially
significantly affect the effective velocities and diffusivities within states
(Subsection~\ref{ssec:homtwomotor}), as well as the effective detachment
rates (Subsection~\ref{subsub:effdet}).  Moreover the magnitude of the jumps
of the cargo tracking variable at detachment (Subsection~\ref{subsub:detach})
and attachment events (Subsubsection~\ref{subsub:attach}) is also sensitive
to the value of the tether spring constant.

These jumps, which perhaps can be associated to cargo flyback~\cite{trybus2013pauses,mallik2018coin,feng2018motor},
can have a nontrivial impact on the effective transport of ensembles of motors.
 The mean cargo jump at detachment of a motor is typically nonzero (Eq.~\eqref{quickformend}),
due to relaxation of the cargo to a new equilibrium balancing the applied
trap force with one tether rather than two, and the preferential detachment
of the leading or trailing motor.  But, at least in our model, the cargo
jump has mean zero upon reattachment of the second motor (Eqs.~\eqref{quickformatstart}
and~\eqref{quickformatend}).  
So  in principle, the motor-cargo complex can have a substantial contribution
to its effective velocity from a ratcheting process in which, from the state
of two attached motors, one detaches, allowing the cargo to rapidly adjust
by thermal diffusion to a statistical equilibrium with the attached motor
while the detached motor even more quickly equilibrates about the cargo,
then the detached motor reattaches to a relaxed configuration (with no net
mean cargo position change), and the two attached motors move again toward
a more strained configuration leading to motor detachment.  This phenomenon
caused a speedup in our model of the motor-cargo complex under an assisting
trap force which in some cases exceeded the maximum single motor velocity.
  
We plan to examine this flyback effect with more biophysical detail in future
work.
   
\subsection{Future Work for Improved Biophysical Fidelity}
Our primary goal in this work has been to set out a mathematical framework
for relating the various physicochemical properties of dissimilar cooperative
motors on their effective transport as a team.  While we have endeavored
to parameterize our models for kinesin-1 and kinesin-2 with experimentally-based
values, we must note a few issues in this parameterization that require further
study before we can meaningfully confront our model predictions to experimental
data on multiple motor transport.  First of all, our parameters from Table~\ref{paramtable}
are all from in vitro measurements.    
As noted as well in McKinley, Athreya, et al~\cite{mckinley2012asymptotic},
the viscosity in cell
is substantially higher than water, and this can make the scale separation
assumption between cargo and attached motor dynamics (small $ \epsilon^{(i)}
$ in Table~\ref{nondtable}) less tenable.  Other biophysical parameters may
also have different values in cell~\cite{kunwar2011mechanical}, though these
are even more difficult to establish than their in vitro counterparts.  Thus,
our focus remains for now on targeting our mathematical framework toward
understanding and interpreting in vitro observations.

At least two parameterization problems require substantial development before
this can be credibly attempted, though.  First of all, to keep focus on the
various
coarse-graining relationships we employed, we adopted in the main text and
in the simulations a simple Hookean spring model for the motor-cargo tether,
with a spring constant obtained from experimental observations of motor-cargo
systems with the tether pulled to its natural extension,  This linear spring
approximation is reasonable for describing small fluctuations of the extended
tether, but does not at all model the tether well when its end-to-end separation
is smaller than its natural extension.  A simple general model used in biophysical
simulations of kinesin is to have a linear restoring force under extension
from a rest length (varying from 40 nm in~\cite{arpag2019kin1kin3} to 80
nm in~\cite{lipowsky2013network} to 110 nm in~\cite{kunwar2010robust,gross2011arranged}),
but no resistance to compression; more complex nonlinear models for extension
have also been considered~\cite{uppulury2013varying,hendricks2009collective,driver2011productive}.
 Using a fully Hookean model with zero rest length and the experimentally
measured linear spring constant (shown in Table~\ref{paramtable}) leads to
an absurd conclusion that the root-mean-square extension of the tether for
kinesin-1 is about 5 nm, when of course it should be more like 70 nm.  The
5 nm is really an estimate for the magnitude of the fluctuations about this
rest length when force on the tether pulls it approximately taut.  During
a given phase of attachment, this neglect of the rest length of the motor-cargo
tether does not necessarily have a strong impact on the calculations -- we
can imagine the cargo is in fact just approximately this rest length behind
the nominal cargo position $ Z(t) $ and the motors would feel generally comparable
forces as in the simple Hookean model with zero rest length which we used.
 The big difference, though, would be on the dynamics of the detached motors,
which should be fluctuating over a distance comparable to the motor-cargo
tether rest length rather than the nominal root-mean-square extension of
the Hookean spring model with zero rest length.  This would have a big impact
on where the detached motors reattach on the microtubule.  Thus, the Hookean
spring model with zero rest length for the motor-cargo tether cannot be expected
to give useful predictions for the transport of actual molecular motors;
we must at least extend it to a nonlinear model with no resistance to compression
below a finite rest length as in~\cite{lipowsky2013network,kunwar2008stepping,kunwar2016anisotropy,arpag2019kin1kin3,gross2011arranged,furuta2013measuring}.
 In Appendix~\ref{app:nonlinspring}, we indicate how nonlinear spring models
for the motor-cargo tether can be handled by our mathematical framework --
the main complication is the jumps of the cargo tracking variable in Section~\ref{sec:effswitch}
become non-Gaussian.  We have here stayed with the purely linear spring model
in the main text to minimize technical complications and more clearly illustrate
the key concepts in the mathematical coarse-graining of the dynamics of a
system of cooperative dissimilar motors.  In future work,
we will adapt this framework to more biophysically accurate nonlinear spring
models for the motor-cargo tether, and examine through this lens various
hypotheses and experimental observations regarding the dynamics of cooperative
molecular motors.

A second parameterization problem is that of the double exponential force-detachment
rate relation~\eqref{demodel}.  Our model is based on a  relation between
detachment rate and instantaneous force, while what is measured~\cite{milic2014kinesin,milic2017intraflagellar,howard2019direction}
is the relation between run length and applied trap force.  We convert the
run length to a detachment rate by dividing by the unloaded velocity, which
is a bit crude but arguably a reasonable rough approximation, but the more
serious concern is equating the dependence on \emph{applied trap} force with
the dependence on \emph{instantaneous} force felt by the motor via its tether
to the cargo.  This leads to particular peculiarities, noted in Section~\ref{sec:sims},
in the absence of trap force, since the force-detachment rate model is discontinuous
at zero force.  So, in our model, the motor at zero trap force is fluctuating
between high detachment rates when the cargo is pulling the motor forward
and low detachment rates when the cargo is pulling the motor backward, leading
to an inappropriately augmented detachment rate.  Rectifying this detachment
rate model requires, in future work, a better statistical inference approach
for a relation between instantaneous force and detachment rate that would
replicate, under our model, the relation between run length and applied trap
force reported in~\cite{milic2014kinesin,milic2017intraflagellar}.  A further
potential improvement would be to incorporate the findings of a recent
study~\cite{howard2019direction}  suggesting a different structure for the
dependence of the detachment
rate against a truly longitudinal applied force on kinesin-1.

\subsection{Future Mathematical Directions}
A mathematical question for future work is to examine whether the coarse-grained
model developed here would change if we started with a discrete stepping
model~\cite{bouzat2016models,reld:rgovm,reld:rsmp,hughes2012kinesins,WangLi:2009,tce:mdbt,abk:mm,driver2010coupling,lipowsky2013network}
for the motor dynamics, rather than the coarse-grained stochastic differential
equation model~\eqref{motors} used here.  This is essentially a question
of how the coarse-graining of a jump process model for the
motor dynamics into a stochastic differential equation interacts with the
coarse-graining procedures developed here.  Perhaps the cargo fluctuations
interacting with the discrete stepping model would give rise to a different
effective dynamics~\eqref{xbar} for the attached motors.

Another question is how our detailed analysis of effective transport of an
ensemble of two cooperative motors can be extended to more general scenarios
of multiple motors attached to a cargo.  The initial coarse-graining steps
over the detached motor and cargo dynamics  in Section~\ref{sec:nond} applied
to an arbitrary number $N$ of cooperative motors, only at the cost of complexity,
but the coarse-graining over the attached motor dynamics in Section~\ref{slowswitch}
relied on the ability to homogenize the attached motor dynamics for the case
$N=2$ via the explicit stationary distribution for an autonomous single-variable
SDE for the separation $ R $ between the motors.  For $ N>2 $, we would have
$N-1 $ coupled degrees of freedom for the internal configuration of the motors,
whose stationary distribution does not seem possible to compute analytically
due to the absence of a potential structure for the drift.  The work of Miles,
Lawley, and Keener~\cite{miles2018analysis}
could readily handle arbitrary numbers of cooperative motors because their
model did not include explicit reference to the spatial configuration of
the motors.  

A complementary extension  to the tug-of-war case of two opposing motors
is problematic because the scale separation assumption between attached motor
dynamics and the detachment process is unlikely to be supported under realistic
models for the detachment rate.  For two cooperative motors, the separation
between them can plausibly be thought to reach a stationary distribution
before they detach, and the nondimensional analysis in Section~\ref{sec:nond}
 provides some quantitative support.  Recall that the predictions from the
asymptotic theory were not so accurate under the double exponential detachment
rate model for a kinesin-2/kinesin-2 ensemble (Table~\ref{resultsde}) under
hindering trap force because the detachment  may actually be governed by
when a fluctuation, perhaps from the initial reattachment configuration,
in the separation of the motors caused the leading motor to feel a higher
force and an exponentially increased detachment rate.  This phenomenon would
be manifest more generally in the case of opposing motors because their separation
would directly increase by their dynamics, only stabilized when the forces
on the motors are substantial enough to slow their rate of separation, but
then the detachment rate would also be substantially increased.  Only if
the stall force for the motors were considerably less than the force scale
of detachment could the dynamics of a pair of attached motors plausibly reach
a stationary distribution before detachment.  But this is certainly not true
for kinesin-1 or kinesin-2 (Table~\ref{paramtable})) and does not appear
relevant for biological motors in general~\cite{mjim:btmme}.
 A fundamentally different mathematical analysis would be required to characterize
the statistics of displacement and detachment from a state of two attached
opposing motors when the dependence of the detachment rate on force varies
strongly over the scale of the stall force of the motors.  

While the model and analysis here considered the case of the cooperative
motors all being associated with a single microtubule, the model and calculations
could still be appropriate for the context of the motors attaching and detaching
from a parallel bundle of microtubules, so long as their spacing is tight
enough that the transverse displacements of the motors could be treated as
having  a negligible effect on the dynamics along the longitudinal direction,
and the cargo dynamics are not much affected by steric interaction with the
microtubule bundles.  These conditions are probably not satisfied in most
biologically relevant contexts, though they could apply to certain engineered
in vitro configurations.  But once the microtubule network is not aligned
with a common polarity, the motor dynamics will be affected by tug-of-war
considerations and the coarse-graining of attached motor dynamics would require
a different approach, as for the case of opposing motors.

\textbf{Acknowledgements:}
This work grew out of conversations between JF and PRK while both were long
term visitors at the Isaac Newton Institute for the program on ``Stochastic
Dynamics in Biology:  Numerical Methods and Applications.''  The authors
also would like to thank Will Hancock for discussions informing the model
development.  PRK would also like to acknowledge early discussions
with Leonid Bogachev, Leonid Koralov, and Yuri Makhnovskii which helped me
map out the general mathematical framework and approaches, while we were
all supported as long-term visitors at the  Zentrum f\"{u}r Interdisziplin\"{a}re
Forschung for the program on ``Stochastic Dynamics: Mathematical Theory and
Applications.''  Here we've managed to work out one of the easier cases we
considered\ldots

\textbf{Funding} The work of JF and PRK are partially supported by National
Institutes of Health grant R01GM122082 and PRK was partially supported by
a grant from the Simons foundation.  The work of JK is partially supported
by an National Science Foundation RTG grant 1344962.

\appendix

\section{Appendix:  A note on nonlinear spring models}\label{app:nonlinspring}
A linear model for representing the tether between the motor and the cargo
is not particularly accurate.  Better theoretical tether models can involve
a model which is Hookean for extension beyond a rest length, but offers no
resistance to compression ~\cite{lipowsky2013network,kunwar2008stepping,kunwar2016anisotropy,arpag2019kin1kin3,gross2011arranged,furuta2013measuring},
a sigmoidal stiffness
dependence on force~\cite{uppulury2013varying}, or a multiple-component model
for
the motor-cargo tether including separate models for the neck linker and
stalk~\cite{hendricks2009collective}.  
We may generalize the averaging results by considering a nonlinear spring
\begin{equation} 
F^{(i)} (r)=\bar F^{(i)}\Phi'^{(i)}(r/L_c^{(i)}) \quad 1 \le i \le N.
\end{equation}
 Here $\Phi^{(i)}(r)$ is a nondimensionalized spring potential,   $L_c^{(i)}$
is
an
appropriate length scale, and $\bar F^{(i)}$ is a characteristic force magnitude
for each motor index $i$.
We define $ \kappa^{(i)} \equiv \bar F^{(i)}/L_c^{(i)} $ as an ``effective''
spring constant of the nonlinear spring; it agrees with the usual spring
constant when the spring force model is purely Hookean, as in the main text.
 
With the same nondimensionalization
as before, the equations of motion become, for 1
$\le i \le N,$
\begin{align}
&d\tilde X^{(i)}(\tilde t) = \left( \epsilon^{(i)} g(s^{(i)} \tilde \kappa^{(i)}(\lambda^{(i)})^{-1}\Phi'^{(i)}(\lambda^{(i)}(\tilde
X^{(i)}(\tilde t)-\tilde
Z(\tilde t))))d\tilde t+\sqrt {\hat \rho^{(i)} \epsilon^{(i)}}dW^{(i)}(\tilde
t)\right)\Atstate{i}(\tilde t)\\&+ \left ( -\left(\Gamma^{(i)}\lambda^{(i)}\right)^{-1}\tilde
\kappa^{(i)}\Phi'^{(i)}\left(\lambda^{(i)}\left(\tilde
X^{(i)}(\tilde t)-\tilde Z(\tilde t)\right)\right)d\tilde t+(\Gamma^{(i)})^{-1/2}dW^{(i)}(\tilde
t)\right)(1-\Atstate{i}(\tilde t)), \quad \nonumber
\\
&d\tilde Z(\tilde t)=\left( \sum_{j = 1}^N (\lambda^{(i)})^{-1}\tilde \kappa^{(i)}\Phi'(\lambda^{(i)}(\tilde
X^{(j)}(\tilde
t)-\tilde Z(\tilde t)))-\tilde{F}_T\right)d\tilde t + d W_{z}(\tilde
t), 
\end{align}
where we have introduced the new nondimensional parameter \begin{equation}
\lambda^{(i)} = \frac{\sqrt{2k_BT/\bar \kappa}}{L_c^{(i)}}.
\end{equation}
which describes the length-scale of thermally induced variations on the tether
relative to the length scale of variation of the tether force law.
Calculations for  the averaged drifts $\bar g^{(i)}$, and thus $G_+$ and
$G_-$,
are similar, but now involve pairing drift functions with non-Gaussian stationary
distributions for unattached motors and cargo (the forms for these equations
are nearly identical to those found in Appendix A in McKinley, Athreya, et
al~\cite{mckinley2012asymptotic}).
 For detachment
jumps, no assumptions of distribution type are made for $p_R(r)$, and therefore
the calculations  for distributions in Subsection~\ref{subsub:detach} only
need
to be adjusted to refer to the mean cargo position under nonlinear tethers.

The random variable $\dMattjump{i}$ describing the
change of position of the cargo tracking variable after motor attachment
may be computed similarly as in Subsection~\ref{subsub:attach}, but it will
no longer be normally distributed or have mean zero in general.
The calculations in Section~\ref{sec:renewal} otherwise go through for a
nonlinear tether model,
 with
only the modified contribution to the moments of the cargo tracking variable
changes
at attachment and detachment jumps.

\section{Appendix:  Derivation of effective diffusion for two motors}\label{app:twomotorapp}

The following derivation for the effective diffusion of the cargo tracking
variable during a state with both  motors attached to the microtubule follows
the multiscale expansion method illustrated  in Pavliotis and Stuart~\cite{pavliotis2008multiscale},
with rigorous exposition in Veretennikov and Pardoux~\cite{pardoux2003poisson}
for the unbounded state
space case relevant to our application.
  Having
computed the effective drift $ \barvtwo $  in Eq.~\eqref{eq:barvtwo} in this
state,
we pass
to a diffusive scaling centered about this mean drift $ \bar t \rightarrow
t/\epsilon^2 $, 
$ M \rightarrow \epsilon (M-\barvtwo t) $, with the internal
configuration variable $ R $ unscaled ($ R \rightarrow R$).  Note in this
appendix, $ \epsilon $ is just a formal small parameter used to push to long
time; it is unrelated to the physically meaningful nondimensional parameters
$ \epsilon^{(i)} $ and $ \bar \epsilon $ in the main text.  Moreover, for
simplicity
for calculations within this appendix, we use the undecorated variables $
t, M, R $ to describe the dynamics under this centered diffusive rescaling,
which read:
\begin{align}
d  M(t)&=  \frac{1}{\varepsilon}\left( G_+(  R(t))-\barvtwo\right)dt+\frac{\sqrt{\rho^{(1)}(\tilde\kappa^{(1)})^2}}2dW^{(1)}(t)+\frac{\sqrt{\rho^{(2)}(\tilde\kappa^{(2)})^2}}2
dW^{(2)}(t) \label{diffeqn1}\\
d  R(t)&=  \frac{1}{\varepsilon^2} G_-(  R(t)) dt+\frac{1}{\varepsilon}\left(\sqrt{\rho^{(1)}}dW^{(1)}(t)-\sqrt{\rho^{(2)}}
dW^{(2)}(t)\right)  \label{diffeqn2}.
\end{align}
The infinitesimal
generator $\mathcal L$ for (\ref{diffeqn1})-(\ref{diffeqn2}) is defined by
its action on a  test function $v = v(m,r)$, with 
\begin{equation} \label{geneqn}
\mathcal Lv(m,r) = \mathbf{h}\cdot \nabla v+\frac 12 \Gamma : \nabla\nabla
v.
\end{equation}
Here  $\mathbf{h}(m,r) = ((G_+(r)-\barvtwo)/\varepsilon, G_-(r)/\varepsilon^2)$
is the drift vector, and $\frac 12 \Gamma$ is the diffusion tensor, where
\begin{align}
\Gamma &= \begin{bmatrix} \frac {\sqrt{\rho^{(1)}(\tilde\kappa^{(1)})^2}}2&
\frac {\sqrt{\rho^{(2)}(\tilde\kappa^{(2)})^2}}2 \\
\frac {\sqrt{\rho^{(1)}}}\varepsilon& -\frac {\sqrt{\rho^{(2)}}}\varepsilon\\
\end{bmatrix}
\begin{bmatrix} \frac {\sqrt{\rho^{(1)}(\tilde\kappa^{(1)})^2}}2& \frac {\sqrt{\rho^{(2)}(\tilde\kappa^{(2)})^2}}2
\\
\frac {\sqrt{\rho^{(1)}}}\varepsilon& -\frac {\sqrt{\rho^{(2)}}}\varepsilon\\
\end{bmatrix}^T \\&=  \begin{bmatrix} \frac{\rho^{(1)}(\tilde\kappa^{(1)})^2+\rho^{(2)}(\tilde\kappa^{(2)})^2}4&
\frac{\rho^{(1)}\tilde\kappa^{(1)}-\rho^{(2)}\tilde\kappa^{(2)}}{2\varepsilon}
\\
\frac{\rho^{(1)}\tilde\kappa^{(1)}-\rho^{(2)}\tilde\kappa^{(2)}}{2\varepsilon}
&\frac{\rho^{(1)}+\rho^{(2)}}{\varepsilon^2}\\
\end{bmatrix}.
\end{align}
We have also used notation for the Frobenius inner product
for matrices, where for matrices $A = (a_{i,j})_{nxm}$ and  $B = (b_{i,j})_{nxm}$,
we define  $A:B = \sum_{i,j} a_{i,j}b_{i,j}$.

We may write out (\ref{geneqn}) explicitly as \begin{align}
\mathcal Lv(m,r) &= \mathbf{h}\cdot \nabla v+\frac 12 \Gamma : \nabla\nabla
v\\
&= \frac{1}{\varepsilon}(G_+(r)-\barvtwo)v_m+\frac{1}{\varepsilon^2}
G_-(r)v_r\\&+\frac 12 \left [ \left(\frac{\rho^{(1)}(\tilde\kappa^{(1)})^2+\rho^{(2)}(\tilde\kappa^{(2)})^2}4\right)v_{mm}+
\frac{\rho^{(1)}\tilde\kappa^{(1)}-\rho^{(2)}\tilde\kappa^{(2)}}{\varepsilon}v_{mr}
+ \left(\frac{\rho^{(1)}+\rho^{(2)}}{\varepsilon^2}\right)v_{rr}\right].
\end{align}
The generator may be decomposed  to match powers of $\varepsilon$ as 
\begin{align}
\mathcal L = \frac{1}{\varepsilon^2}\mathcal L_0+ \frac 1{\varepsilon}\mathcal
L_1+ \mathcal L_2,
\end{align}
with
\begin{align}
\mathcal L_0 &= G_-(r)\partial_r +\frac {\rho^{(1)}+\rho^{(2)}}2\partial_{rr},\\
\mathcal L_1 &= \frac{\rho^{(1)}\tilde\kappa^{(1)}-\rho^{(2)}\tilde\kappa^{(2)}}{2}\partial_{mr}+\left(G_+(r)-\barvtwo\right)\partial_m,\\
\mathcal L_2 &=  \left(\frac{\rho^{(1)}(\tilde\kappa^{(1)})^2+\rho^{(2)}(\tilde\kappa^{(2)})^2}8\right)\partial_{mm}.
\end{align}
 Assuming a multiscale solution $v = v_0+\varepsilon v_1+\varepsilon^2 v_2
+\dots$ for the backward Kolmogorov equation 
\begin{equation}
\frac{\partial v}{\partial t} = \mathcal Lv,
\end{equation}
we  match powers of orders $1/\varepsilon^2, 1/\varepsilon,$ and $1$ to obtain
\begin{align}
\mathcal L_0 v_0 &= 0,\\
-\mathcal L_0 v_1 &= \mathcal L_1 v_0,\\
-\mathcal L_0 v_2 &= -\frac{\partial v_0}{\partial t}+ \mathcal L_1 v_1+
\mathcal
L_2 v_0. \label{oper3}
\end{align}
The first equation implies that $v_0$ is only a function of $m$ and $t$.
From here, the second equation may be simplified to 
\begin{equation}
-\mathcal L_0 v_1 =\left(G_+(r)-\barvtwo\right)\partial_{m}v_0(m,t).
\end{equation}
As the operator $\mathcal L_0$ only depends on  $r$, we may express
$v_1$ as 
\begin{equation}
v_1(m,r,t) = \chi(r)\partial_{m}v_0(m,t).
\end{equation}
Proceeding to the third equation of the asymptotic expansion,  the Fredholm
alternative states that for (\ref{oper3}) to have a solution, its right hand
side must be orthogonal to $p_R(r)$, or
\begin{align}
\frac{\partial v_0}{\partial t} &= \int_\mathbb R p_R(r)\mathcal L_2 v_0(m,t)dr+\int_\mathbb
R p_R(r)\mathcal L_1(\chi(r)\partial_m v_0(m,t))dr\\
 &: = I_1+I_2.
\end{align}

We look at each integral in turn.  First,
\begin{align}
  I_1 &= \int_\mathbb R p_R(r) \mathcal L_2v_0(m,t) dr\\&= \int_\mathbb
R p_R(r)\left[\left(G_+(r)-\barvtwo\right)\partial_m v_0
(m,t)+\left(\frac{\rho^{(1)}(\tilde\kappa^{(1)})^2+\rho^{(2)}(\tilde\kappa^{(2)})^2}8\right)\partial_{mm}v_0(m,t)\right]dr\\
  &= \left(\frac{\rho^{(1)}(\tilde\kappa^{(1)})^2+\rho^{(2)}(\tilde\kappa^{(2)})^2}8\right)\partial_{mm}v_0(m,t).
  \end{align} 
The second integral may be broken up further, as
\begin{align}
I_2 &= \int_\mathbb R p_R(r)\mathcal L_1(\chi(r)\partial_m v_0(m,t))dr\\
 &= \int_\mathbb R p_R(r)\Big[\left(\frac{\rho^{(1)}\kappa^{(1)}-\rho^{(2)}\kappa^{(2)}}{2}\right)\partial_{mr}(\chi(r)\partial_m
v_0(m,t))\\&+\left(G_+(r)-\barvtwo\right)\partial_m(\chi(r)\partial_m
v_0(m,t))\Big]dr \\
 &:= I_3+I_4.
\end{align}
The first part satisfies
\begin{align}
I_3 &= \int_\mathbb R p_R(r)\left(\frac{\rho^{(1)}\tilde\kappa^{(1)}-\rho^{(2)}\tilde\kappa^{(2)}}{2}\right)\partial_{mr}(\chi(r)\partial_m
v_0(m,t))dr\\
 &= \left(\left(\frac{\rho^{(1)}\tilde\kappa^{(1)}-\rho^{(2)}\tilde\kappa^{(2)}}{2}\right)\int_\mathbb
R p_R(r)\partial_{r}\chi(r)dr\right)\partial_{mm} v_0(m,t) \label{a1calc}\\
&:= A_1\partial_{mm} v_0(m,t).
\end{align}
Finally, we have
\begin{align}
I_4 &=  \int_\mathbb R p_R(r)\left[\left(G_+(r)-\barvtwo\right)\partial_m(\chi(r)\partial_m
v_0(m,t))\right]dr \\
 &=  \int_\mathbb R p_R(r)\left[\left(G_+(r)-\barvtwo\right)\chi(r)\right]dr\partial_{mm}
v_0(m,t)\\
&:= A_2\partial_{mm} v_0(m,t). \label{eq:a2}
\end{align}

The closed form equation for $v_0(m,t)$ is thus given by

\begin{equation}
 \frac{\partial v_0}{\partial t} =  \frac 12\left(\frac{\rho^{(1)}(\tilde\kappa^{(1)})^2+\rho^{(2)}(\tilde\kappa^{(2)})^2}4+2A_1+2A_2\right)\partial_{mm}v_0(m,t),
\end{equation}
and is the backward Kolmogorov equation for the SDE
\begin{align}\label{homtwomotor}
d M(t) &=\sqrt{\frac{\rho^{(1)}(\tilde\kappa^{(1)})^2+\rho^{(2)}(\tilde\kappa^{(2)})^2}4+2A_1+2A_2}dW(t)\\
&\equiv  \sqrt{2D^{(1,2)} }dW(t),
\end{align}
where $W(t)$ is a standard Brownian motion.

Now we compute explicit expressions for constants $A_1$ and $A_2$. 
This involves solving the cell problem
for  $\chi$, given by
\begin{align}
 -G_-(r)\chi'(r)-\left(\frac{\rho^{(1)}+\rho^{(2)}}2\right) \chi''(r) &=
\tilde g_+ \label{cell1}(r),\\ \quad \int_{\mathbb R}\chi(r)p_R(r)dr
&= 0,
\end{align}  
where we define $\tilde g_+(r) =G_+(r) -\barvtwo$.
If we rewrite (\ref{cell1}), using an integration factor $\mu(r)$, as
\begin{equation}\label{intfact}
[\mu(r)\chi'(r)]' = -\mu(r)\tilde g_+(r)\left(\frac{2}{\rho^{(1)}+\rho^{(2)}}\right),
\end{equation}
then it is straightforward to show that   $\mu(r) $ is in fact equal to
the stationary distribution $p_R(r)$.

Integrating out (\ref{intfact}) leaves us with\begin{equation}
\chi'(r) =- \int_{-\infty}^r \tilde g_+(r^{\prime})\left(\frac{2}{\rho^{(1)}+\rho^{(2)}}\right)p_R(r^{\prime})dr^{\prime}/p_R(r)+C/p_R(r)
\end{equation}
for some unknown integration constant $C.$ 
By the subexponential growth requirement on $\chi$ and $\chi'$~\cite{pardoux2003poisson},
it follows that $C=
0$.
From (\ref{a1calc}), \begin{align}
A_1 &= \left(\frac{\rho^{(2)}\tilde\kappa^{(2)}-\rho^{(1)}\tilde\kappa^{(1)}}{\rho^{(1)}+\rho^{(2)}}\right)\int_\mathbb
R \int_{-\infty}^r \tilde g_+(r^{\prime})p_R(r^{\prime})dr^{\prime} dr\\
&=  \left(\frac{\rho^{(1)}\tilde\kappa^{(1)}-\rho^{(2)}\tilde\kappa^{(2)}}{\rho^{(1)}+\rho^{(2)}}\right)\int_\mathbb
R  \tilde g_+(r)p_R(r)\cdot rdr. 
\end{align}  
The last equality used integration by parts, in which the boundary term vanishes
under the assumption that $p_R(r)= o(1/r^2)$ as $r \rightarrow \pm \infty$.
The calculation for $A_2$~\eqref{eq:a2} follows  from integration by parts,
with \begin{align}
 A_2 &= \int_{-\infty}^\infty \chi  (r)\tilde g_+ (r)p_R(r)dr =- \int_{-\infty}^\infty
\chi(r) (\mathcal L_0 \chi(r)) p_R(r)dr\\
 &=- \int_{-\infty}^\infty \chi(r) \left( G_-(r)\chi' (r)+\left(\frac{\rho^{(1)}+\rho^{(2)}}2\right)\chi''
(r)\right)p_R(r)dr\\ 
 &= \int_{-\infty}^\infty \left(\frac{\rho^{(1)}+\rho^{(2)}}2\right)\chi'(r)^2p_R(r)dr\\
  &= \int_{-\infty}^\infty \left(\frac{2}{\rho^{(1)}+\rho^{(2)}}\right)\left(\int_{-\infty}^r
\tilde g_+ (r^{\prime}) p_R(r^{\prime} )dr^{\prime} \right)^2\frac{1}{p_R(r)}dr.
    \end{align}


\bibliographystyle{siam}    
\bibliography{motorreference}

\end{document}